\documentclass[aps,prc,twocolumn,nofootinbib,superscriptaddress]{revtex4-1}

\usepackage{bm}        
\usepackage{xcolor}

\usepackage{dcolumn}   
\usepackage{multirow}

\usepackage{amsmath}
\usepackage{amsfonts}
\usepackage{amssymb}
\usepackage{makeidx}
\usepackage{graphicx}
\usepackage{anysize}
\usepackage{hyperref}
\usepackage{slashed}
 \usepackage{bbm}

\usepackage{xcolor}

\usepackage{ulem} 

\usepackage{colortbl} 

\newenvironment{mymathbox}
{\par\smallskip\centering\begin{lrbox}{0}%
\begin{minipage}[c]{0.8\textwidth}}
{\end{minipage}\end{lrbox}%
\framebox[0.9\textwidth]{\usebox{0}}%
\par\medskip
\ignorespacesafterend}
\newcommand{\bb}{\begin{mymathbox}}
\newcommand{\eb}{\end{mymathbox}}

\newcommand{\be}{\begin{equation}}
\newcommand{\ee}{\end{equation}}
\newcommand{\ba}{\begin{eqnarray}}
\newcommand{\ea}{\end{eqnarray}}

\newcommand{\nk}{{\bf      k}}
\newcommand{\np}{{\bf      p}}
\newcommand{\nq}{{\bf      q}}

\newcommand{\npsi}{{\bf \npsi}}

\newcommand{\bma}{\begin{pmatrix}}
\newcommand{\ema}{\end{pmatrix}}

\begin{document}

\title{Electron-induced single-pion production to constrain the neutron structure in $^{40}$Ar. A proof of concept.}

\author{J. García-Marcos}
\email{javier31@ucm.es}
\affiliation{Grupo de Física Nuclear, Departamento de Estructura de la Materia, Física Térmica y Electrónica, Facultad de Ciencias Físicas, Universidad Complutense de Madrid and IPARCOS, CEI Moncloa, Madrid 28040, Spain}
\affiliation{Department of Physics and Astronomy, Ghent University, B-9000 Gent, Belgium}
\author{M. Hooft}
\affiliation{Department of Physics and Astronomy, Ghent University, B-9000 Gent, Belgium}
\author{T. Franco-Munoz}
\email{Current affiliation: Department of Physics and Astronomy, Ghent University, B-9000 Gent, Belgium. Tania.FrancoMunoz@ugent.be} 
\affiliation{Grupo de Física Nuclear, Departamento de Estructura de la Materia, Física Térmica y Electrónica, Facultad de Ciencias Físicas, Universidad Complutense de Madrid and IPARCOS, CEI Moncloa, Madrid 28040, Spain}
\author{N. Jachowicz}
\email{natalie.jachowicz@ugent.be}
\affiliation{Department of Physics and Astronomy, Ghent University, B-9000 Gent, Belgium}
\author{J.M.~Udías}
\email{jmudiasm@ucm.es}
\affiliation{Grupo de Física Nuclear, Departamento de Estructura de la Materia, Física Térmica y Electrónica, Facultad de Ciencias Físicas, Universidad Complutense de Madrid and IPARCOS, CEI Moncloa, Madrid 28040, Spain}
\author{K. Niewczas}
\affiliation{Department of Physics and Astronomy, Ghent University, B-9000 Gent, Belgium}
\author{A. Nikolakopoulos}
\email{anikolak@uw.edu}
\affiliation{University of Washington, Seattle, WA 98195, USA}
\author{R. González-Jiménez}
\email{raugj@us.es}
\affiliation{Departamento de Física Atómica, Molecular y Nuclear, Universidad de Sevilla, 41080 Sevilla, Spain}



\date{\today}

\begin{abstract}
We study electron-induced single-pion production as a way to constrain the neutron structure of $^{40}$Ar, information that is necessary for neutrino experiments using argon detectors. The proposed experimental signal consists in detecting in coincidence the scattered electron, a proton and a $\pi^-$. We performed simulations compatible with the experimental conditions of the MAMI (University of Mainz) and CLAS (Jefferson Lab) facilities. We have computed cross sections and evaluated the main backgrounds. 
MAMI is a three-spectrometer system with extremely good energy resolution and small acceptances. We found that, by choosing specific values of the final particle's momenta, the shell structure can be well resolved, with negligible background contributions.
CLAS is a large solid angle detector with poorer energy resolution. In both cases, the background can be kept under control by performing cuts in missing energy and missing momentum; however, in CLAS, the shell structure cannot be resolved due to the energy resolution.
We conclude that MAMI is particularly appropriate for the proposed experiment, whereas CLAS is better suited for other studies.
\end{abstract}

\maketitle


\section{Introduction}

Liquid argon is the detector material in some present and future neutrino experiments: MicroBooNE~\cite{MicroBooNEweb}, SBND~\cite{SBNDweb}, ICARUS~\cite{ICARUSweb} and DUNE~\cite{DUNEweb}.
The uncertainties derived from the modeling of the neutrino-nucleus cross section, together with the normalization of the neutrino flux, are the main contribution to systematic uncertainties in accelerator-based neutrino oscillation experiments, and will compete with the statistical uncertainties in the new generation of oscillation experiments~\cite{Alvarez-Ruso25,T2K23}.
 
The neutrino energy is a key parameter in oscillation analyses. 
Since neutrino beams are not monoenergetic, for each event in the detector, corresponding to a configuration of particles, one needs to include the probability distribution of the neutrino energies consistent with this event. 
This process, known as neutrino energy reconstruction, relies on the knowledge of the neutrino-nucleus cross section which is therefore a source of systematic uncertainties~\cite{Mahn18, Martini12, Nieves12a, Lalakulich12a, Nikolakopoulos18a}.
Below the deep-inelastic region, neutrino-nucleus interactions may be described by considering nucleons as degrees of freedom: the neutrino interacts with protons or neutrons bound in the nucleus. 
The single nucleon (hole) spectral function is of particular importance. 
The spectral function provides the joint probability of removing a nucleon with certain momentum while leaving the residual system with certain excitation energy and can be defined and computed from the ground state of a many-body system~\cite{ManyBodyTheoryExposed}. How the spectral function enters into the computation of the interaction rate depends on the possible approximations made~\cite{Boffi93}. 
In any case, the spectral function determines the spectrum of excitation energy and momentum of the residual system in nucleon-removal reactions.
Thereby, it is crucial to predict the correct energy balance in neutrino interactions, and subsequently, to minimize errors in the reconstruction of the neutrino energy~\cite{Chakrani24,Baudis24}.

Recently, the E12-14-012 experiment~\cite{Gu21}, performed in Jefferson Lab Hall A~\cite{Jlabwebsite}, studied the spectral function of $^{40}$Ar.
Results for the proton spectral function, obtained from the analysis of $^{40}$Ar$(e,e'p)$ data, were reported in \cite{Jiang22}.
Because the neutrino and antineutrino couple to neutrons and protons respectively, both neutron and proton spectral functions are needed. 
Although it would be possible to study the reaction $^{40}$Ar$(e,e'n)$, there are obvious complications related to neutron detection. 
Thus, to extract information on the neutron spectral function, the E12-14-012 experiment used $^{48}$Ti as target~\cite{Jiang23}. The idea is based on the hypothesis that the proton spectral function of $^{48}$Ti approximately coincides with the neutron spectral function of $^{40}$Ar. This is, up to a few MeV, supported by some nuclear theory predictions, e.g., the ab initio self-consistent Green's function approach~\cite{Barbieri19} and relativistic and non-relativistic independent-particle shell-model approaches~\cite{VanDessel23,Franco-Patino24}. 

The energy levels of neutrons in $^{40}$Ar are relatively well known, at least for the outer shells. For example, the most external shell is known to be at 9.87 MeV~\cite{nuclear-masses}, which corresponds to the neutron separation energy. 
There is also information from experiments on other reactions such as $^{40}$Ar$(p,d)$, $^{40}$Ar$(d,^3\text{H})$, or $^{40}$Ar$(^3\text{He},\alpha)$ \cite{NuclearDataSheets_A=39}, where the reaction is governed by the strong interaction; therefore, the cross sections are larger. However, these experiments did not provide detailed information about momentum distribution or sufficient missing energy resolution.

Single-pion production is an alternative way to study the nuclear spectral function of neutrons in argon, which would allow one to obtain the $E_m$-$p_m$ distribution.
As an example, Fig.~1 of~\cite{MacKenzie96} shows the experimental missing energy spectrum obtained in the reaction $^{12}$C$(\gamma,\pi^-n)$. 
The two main peaks of protons in carbon are nicely identified, with a precision similar to that of other $^{12}$C$(e,e'p)$ experiments under equivalent kinematical conditions~\cite{Garino92,Holtrop98,Dutta03}.

In this work, we investigate the use of a triple coincidence experiment $^{40}$Ar$(e,e'p\pi^-)$, as a way to constrain the neutron spectral function of $^{40}$Ar. 

This reaction was previously studied for $^{12}$C in MAMI~\cite{Blomqvist98}. However, to our knowledge, the only published results for the $^{12}$C$(e,e'p\pi^-)$ reaction is the spectrum of Fig.~32 in Ref.~\cite{Blomqvist98} where the main peaks corresponding to single-neutron removal are nicely identified despite the relatively low statistics.

In quasifree kinematics, the main contribution to this channel is the interaction with a bound neutron
\ba
  e + n \longrightarrow e' + p + \pi^-\,.\label{signal}
\ea
Therefore, the $^{40}$Ar neutron structure is probed directly, instead of through a proxy nucleus like $^{48}$Ti.
The clear advantage of using this reaction is that the detected particles are all charged, hence avoiding the complications due to neutron detection.

The main drawbacks are the low statistics inherent in a triple coincidence experiment, and that the modeling of single-pion production (SPP) is theoretically more challenging than quasielastic nucleon knockout.
The latter should actually be understood as an opportunity to improve our understanding of nuclear effects in SPP in the Delta region.
This is similarly important for neutrino oscillation experiments~\cite{Alvarez-Ruso18}, in particular for experiments that use few-GeV neutrino fluxes.

We stress that there is no cross section data for electron-induced incoherent pion-production on the nucleus in the Delta region. The data of Ref.~\cite{Clasie07} are at relatively high-$Q^2$ and very high pion momentum (above 2.5 GeV), and focused on nuclear transparency, so only cross section ratios were reported.
Some data exist for light nuclei, $^4$He~\cite{Steenbakkers05} and deuteron~\cite{Gilman90}. 
There are data on coherent pion production induced by photons~\cite{Gothe95,Krusche02} and electrons~\cite{Drechsel99,Sambeek97}. 
Datasets on photon-induced exclusive incoherent SPP exist, and are useful to study pion FSI~\cite{Bosted80,MacKenzie96,vanUden98,Hicks00,Krusche04}; however, none of them uses argon as target. 
In summary, the data from the experiments proposed here would be the first of their kind and would represent a unique opportunity to study both the pion FSI in $^{40}$Ar and its shell structure.

We analyze two possible experimental scenarios compatible with the capabilities of the currently active laboratories CLAS@JLab~\cite{CLASwebsite}, and MAMI@Mainz~\cite{MAMIwebsite}. 

This work should be understood as {\it a proof of concept}. Rather than providing the most advanced or comprehensive theoretical models, we aim to present an idea and demonstrate its feasibility, extracting robust conclusions based on kinematics and subsequently, minimizing model dependency.

\section{Kinematics}\label{sec:kin}

\begin{figure}[ht]
\centering  
\includegraphics[width=0.50\textwidth,angle=0]{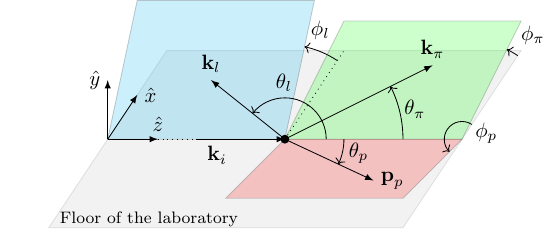}
\caption{Definition of the the reference frame. The $\hat{z}$-axis is defined by the electron beam. The $\hat{x}$-$\hat{z}$ plane coincides with the floor of the laboratory. The scattering angles ($\theta_l$, $\theta_p$ and $\theta_\pi$) are defined on this plane. Azimuthal angles are the angles with respect to the $\hat{x}$-$\hat{z}$ plane.}
\label{fig:planes}
\end{figure}

The reaction $^{40}$Ar$(e,e'p\pi^-)$ is considered. 
The reference frame used in this work is sketched in Fig.~\ref{fig:planes}. 

The missing energy is defined as
\ba\label{Eq:Em}
  E_m = \omega - T_p - E_\pi - T_{R}\,,
\ea
where $\omega=E_i-E_l$ is the energy transferred by the electron to the nuclear system, $E_\pi$ is the energy of the detected pion, $T_p$ stands for the kinetic energy of the detected proton and $T_R$ stands for the kinetic energy of the residual system.
The missing energy measures the excitation energy of the residual system when a nucleon is removed. 

We define the missing momentum, $\np_m$, as the opposite of the residual-system momentum: 
\ba\label{Eq:pm}
  \np_m = \np_p + \nk_\pi - \nq\,,
\ea
where $\nq=\nk_i-\nk_l$ is the momentum transfer.
The residual system has mass
\begin{equation}
    M_R = M_A - M_N + E_m,
\end{equation}
and energy
\ba
    E_R = \sqrt{p_m^2 + M_R^2} = M_A + \omega - E_p -E_\pi.
\ea
We are interested in the distribution of missing energy and momentum for small $E_m < 70~\mathrm{MeV}$ and $p_m < 350~\mathrm{MeV}$. These probe the single-particle hole spectral function: the probability of finding a bound nucleon with given energy and momentum.

We perform simulations of the SPP signal and the main background to show the experimentally attainable resolution of the observable $E_m$ spectrum.
We do not need to consider multi-pion production as a possible background; while it can mimic the $p \pi^-$ signal, the missing energy is always larger than the pion mass $E_m > m_\pi$, and hence it does not contribute to the kinematic region of interest.

The background we consider are the different channels in SPP and single-nucleon knockout, which, through rescattering in the nucleus, can contribute to the experimental signal.
We will show explicitly that these backgrounds can be mostly eliminated by suitable selection of kinematics. 
The selected kinematics are essentially quasifree kinematics, i.e. the limits of Eqs.~(\ref{Eq:Em}-\ref{Eq:pm}) for small $E_m$ and $p_m$.
These limits imply strong correlation between the energies and scattering angles of the pion and proton. Signal events, from direct $p\pi^- $ production, will cluster around these values, while rescattering causes the kinematics to deviate. This allows selecting kinematics that maximize the signal and practically eliminate the background.

\section{Theoretical models}\label{theomodels}

\subsection{Single-pion production model}
We use the Ghent-Hybrid model, published in~\cite{Gonzalez-Jimenez17, Nikolakopoulos23}. It consists of a combination of a low-energy model and a high-energy, Regge-based, model. For the kinematics studied here, only the low-energy component matters, which is essentially identical to that of~\cite{Hernandez07}. Single-pion production happens through the Delta resonance, other higher-mass resonances and a non-resonant background from a $\chi$PT model. All contributions are added coherently, at the amplitude level. 
The model has been compared to electro- and photo-production data and other state-of-the-art models in~\cite{Nikolakopoulos23, NikolakopoulosPhD}.\\

\subsection{Nuclear model}
We work in the relativistic plane-wave impulse approximation (RPWIA) approach. The initial state is described by an independent-particle shell model (IPSM), the relativistic mean-field model (RMF)~\cite{Serot97,computational-nuclear-physics-book}, with the set of parameters of Ref.~\cite{NLSH}. 
In Table~\ref{table:levels}. we show the IPSM predictions for the energy eigenvalues and occupations of the initial-state nuclear levels. 

The knockout nucleon and pion are described using relativistic plane waves.

\begin{table}
    \centering
    \begin{tabular}{c c c c c c c c c c c c c}
        \hline
        \hline
        \arrayrulecolor{white}\hline
        \hline
     & Level & & & $(BE)_n$ [MeV] & & $N_n$ & & & $(BE)_p$ [MeV] & & $N_p$\\
        \arrayrulecolor{white}\hline
        \hline
        \arrayrulecolor{black}\hline
        \arrayrulecolor{white}\hline
        \hline
     & $1s_{1/2}$ & & & 54.56 & & 2 & & & 47.42 & & 2 & \\ 
     & $1p_{3/2}$ & & & 38.74 & & 4 & & & 32.31 & & 4 & \\  
     & $1p_{1/2}$ & & & 34.20 & & 2 & & & 27.72 & & 2 & \\
     & $1d_{5/2}$ & & & 23.09 & & 6 & & & 17.14 & & 6 & \\
     & $1d_{3/2}$ & & & 15.95 & & 4 & & & 10.02 & & 2 & \\
     & $2s_{1/2}$ & & & 16.11 & & 2 & & &  9.86 & & 2 & \\
     & $1f_{7/2}$ & & & 8.27  & & 2 & & &       & &  & \\
    \arrayrulecolor{black}\hline
     \hline
    \end{tabular}
    \caption{Predicted binding energy $(BE)_i$ and number of nucleons ($N_i$) with the RMF model~\cite{NLSH}. Second and third columns correspond to neutrons ($i\equiv n$), fourth and fifth to protons ($i\equiv p$).}
    \label{table:levels}
\end{table}

In this work, the energy spectrum of the nuclear levels, which is a set of Dirac deltas in the IPSM, is replaced by Gaussian distributions with very narrow widths ($\sigma=0.5$ MeV). 
The occupation of all levels is reduced to 80\% of the {\it full} occupation predicted by the IPSM.
In a full experimental analysis, the goal would be to determine empirically widths and occupation numbers of such states. 
The simple spectrum adopted here serves the purpose of clearly illustrating the attainable resolution of an experimental measurement, while crucially retaining realistic momentum distributions and hence scattering kinematics.

In addition to the mean field contribution, we consider that 20\% of nucleons are in SRC pairs, which is consistent with recent experiments~\cite{Duer18,Egiyan06}. 
The resulting SRC background is modeled as a function starting at the $E_m$ value corresponding to the two-nucleon knockout threshold and extending to very high-$E_m$ and high-$p_m$ regions. 
The shape, in terms of $E_m$ and $p_m$, is taken from both $(e,e'p)$ experimental data and theory~\cite{Benhar08}. 
This is the approach of~\cite{Benhar08,Gonzalez-Jimenez22,Franco-Patino22,Franco-Patino24} used to model QE scattering. 
With this semi-phenomenological approach, one includes the interaction with a single nucleon in an SRC pair. The kinematics of the leading nucleon that is knocked out are explicitly described, while all possible configurations of the second nucleon are integrated out. 
It would be desirable to have a microscopic approach for these $NN'$ and $NN'\pi$ processes; however, this represents our best available approach at present to estimate this contribution.

Again, our goal in this work is to study the possibilities of the proposed experiments to resolve the shell structure of neutrons in $^{40}$Ar. Therefore, a more complicated spectral function model is not necessary at this point.\\

\subsection{Final-state interactions}

Rescattering of nucleons and pions with the residual nucleus can occur if energy-momentum conservation allows it.
We model this using the \textsc{NuWro} intranuclear cascade (INC) model~\cite{Golan12, Niewczas19}, using the same approach as in~\cite{Nikolakopoulos2022, Nikolakopoulos2024}, which consists in propagating the hadrons provided by an `external' model, in our case the RPWIA, through the nucleus. 

The \textsc{NuWro} INC propagates hadrons in space using a classical Monte Carlo method based on the probability of particle interactions in the medium.
During this propagation, particles may interact with the nucleons of the residual system. The simulation explicitly includes (among others) the knockout of additional nucleons, production of pions in secondary $NN$ interactions, pion-nucleon rescattering, and pion absorption~\cite{Golan12, Niewczas19}.
All these processes may provide contributions to the observable signal.

The first main effect of the INC is to reduce the signal.
A $p \pi^-$ pair may undergo pion absorption or charge-exchange (CEX), thereby removing it from the signal.
The pion or proton may knock out additional nucleons or produce pions, which removes the event from the signal by increasing the missing energy $E_m$.
This reduction is often quantified by nuclear transparency, i.e., the ratio of observed events in quasifree kinematics with respect to a calculation that neglects FSI~\cite{Niewczas19}.

On the other hand, low energy nucleons might be knocked out, thereby increasing the missing energy by an amount of the order of the nucleon separation energy and some kinetic energy, but the event could still contribute to the signal region $E_m < 70~\mathrm{MeV}$.
This is a rescattering background, which is explicitly modeled by performing simulations over the full phase space of the hadrons. 

Another background that is important in this case is CEX. In addition to removing events from the signal, a propagating $\pi^0$ or $\pi^+$ can undergo CEX into the $\pi^-$ channel.
Again, to include these backgrounds we perform a simulation of all pion production channels $n\pi^0, p\pi^0, p\pi^-, n \pi^+$ over the full hadronic kinematic phase space.
In addition, we include the single proton and neutron knockout channels, which may contribute to the signal through secondary pion production.

In conclusion, we include all SPP channels and QE scattering in our simulation to determine the signal and backgrounds. As mentioned earlier, multi-pion production does not contribute to the region of interest since it implies $E_m > m_\pi$. Other contributions such as two particle-two hole induced by meson exchanged currents~\cite{VanCuyck17, Megias15}, which could potentially create a pion due to rescattering are not included. However, this contribution is expected to have a similar (or smaller) effect than contributions from QE scattering due to the fact that the two knocked out nucleons must share the energy and momentum transferred.

        \begin{table*}
        \begin{tabular}{c l c c c c c c c c c c c c c l c }
        \hline
        \hline
        \arrayrulecolor{white}\hline
        \hline
        & Spectrometer Configuration & & & & Spec. A & & & & Spec. B & & & & Spec. C &\\
        \arrayrulecolor{white}\hline
        \hline
        \arrayrulecolor{black}\hline
        \arrayrulecolor{white}\hline
        \hline
        & Maximum momentum [MeV] & & & & 735  & & & &  870 & & & & 551\\ 
        & Scattering angle range [deg] & & & & 18$-$160 & & & & 7$-$62  & & & & 18$-$160\\   
        & Out-of-plane capability [deg] & & & & - & & & & 0$-$10 & & & & -\\
        & Momentum resolution & & & & $\leq 10^{-4}$  & & & &  $\leq 10^{-4}$ & & & & $\leq 10^{-4}$\\ 
        & Angular resolution at target [mrad] & & & & $\leq 3$ & & & & $\leq 3$  & & & & $\leq 3$\\  
        \arrayrulecolor{black}\hline
        \hline
        \end{tabular}
        \caption{Key parameters of the three spectrometers at MAMI. In addition, Spec. A and C rotate on the left and right side respectively, relative to the incoming beam. Spec. B can rotate on the right side. Minimum angle between A and B is 25 deg, and 30 deg between B and C. Angles less than 24 deg for A respect to the outgoing beam, and less than 15 for B should be avoided. Values and information taken from~\cite{Blomqvist98}.
}
        \label{table:resolution}
        \end{table*}

\section{Results}

The goal of this work is to study the feasibility of a $^{40}$Ar($e,e'p\pi^-$) triple coincidence experiment at MAMI and CLAS facilities. To do so, realistic estimates of the backgrounds are essential. The simulation workflow is as follows:
\begin{itemize}
    \item[1.] The events are generated for each channel that is considered: the four SPP channels and the single proton and single neutron knockout.
    \item[2.] The hadrons produced in those events are propagated through the nucleus using the \textsc{NuWro} INC.
    \item[3.]  We select those events that have ``at least one proton and one negative pion'' in the final state, so it potentially can contribute to the signal if both are detected in coincidence together with the scattered electron.
    \item[4.] We perform the analysis taking into account cuts in phase-space and the capabilities of each facility, and compute the cross sections. 
\end{itemize}
Proceeding this way, we expect to obtain robust conclusions about this potential experiment that are based on kinematics, therefore, reducing the model dependency. In other words, we do not try to predict or reproduce the cross section in terms of $E_m$, but to get reliable estimates of the backgrounds at each facility.

\subsection{MAMI}

MAMI is a three-spectrometer facility for electron beams run by the Institute for Nuclear Physics of the University of Mainz~\cite{MAMIwebsite}. The spectrometers have small acceptances and extremely good momentum and angular resolution, summarized in Table~{\ref{table:resolution}}, together with some general considerations regarding the facility.

We perform our study under some plausible experimental conditions.

The acceptances of the spectrometers~\cite{Blomqvist98} and the configuration chosen for our simulation are summarized in Table \ref{table:MAMI}. The choice of kinematics will be justified throughout this section.

        \begin{table*}[ht]
        \begin{tabular}{c c c l c c c c c c c c c c c c c c c c}
        \hline
        \hline
        \arrayrulecolor{white}\hline
        \hline
        & Spec. & & Particle & & & & $p$ [MeV] & $\Delta p$ [MeV] & & & & $\theta$ [deg] & $\Delta\theta$ [deg] & & & & $\phi$ [deg] & $\Delta\phi$ [deg] & \\
        \arrayrulecolor{white}\hline
        \hline
        \arrayrulecolor{black}\hline
        \arrayrulecolor{white}\hline
        \hline
        & A & & Proton & & & & 510 & 10\% & & & &  47.00 & 7.73 & & & &  0 & 4 \\ 
        & B & & Lepton & & & & 505 & 7.5\% & & & &  30.00 & 1.15 & & & &  180 & 4 &  \\  
        & C & & Pion & & & & 130 & 12.5\% & & & &  80.00 & 7.73 & & & & 180 & 4 \\
        \arrayrulecolor{black}\hline
        \hline
        \end{tabular}
        \caption{Selected configuration of the three spectrometers at MAMI. Acceptances are taken from~\cite{Blomqvist98}.
        }
        \label{table:MAMI}
        \end{table*}

We start studying an inclusive sample, where the full phase space of hadrons in the final state is included in the experiment, while the beam energy and lepton scattering angle are fixed. 
This inclusive cross section is useful to calibrate the theory and experiment.

The selection of the lepton kinematics should be carefully made so that the SPP cross section is sufficiently large and, at the same time, the theory describing the process is robust, that is, around the Delta peak.  

\begin{figure}[h]
\centering
\includegraphics[width=.49\textwidth,angle=0]{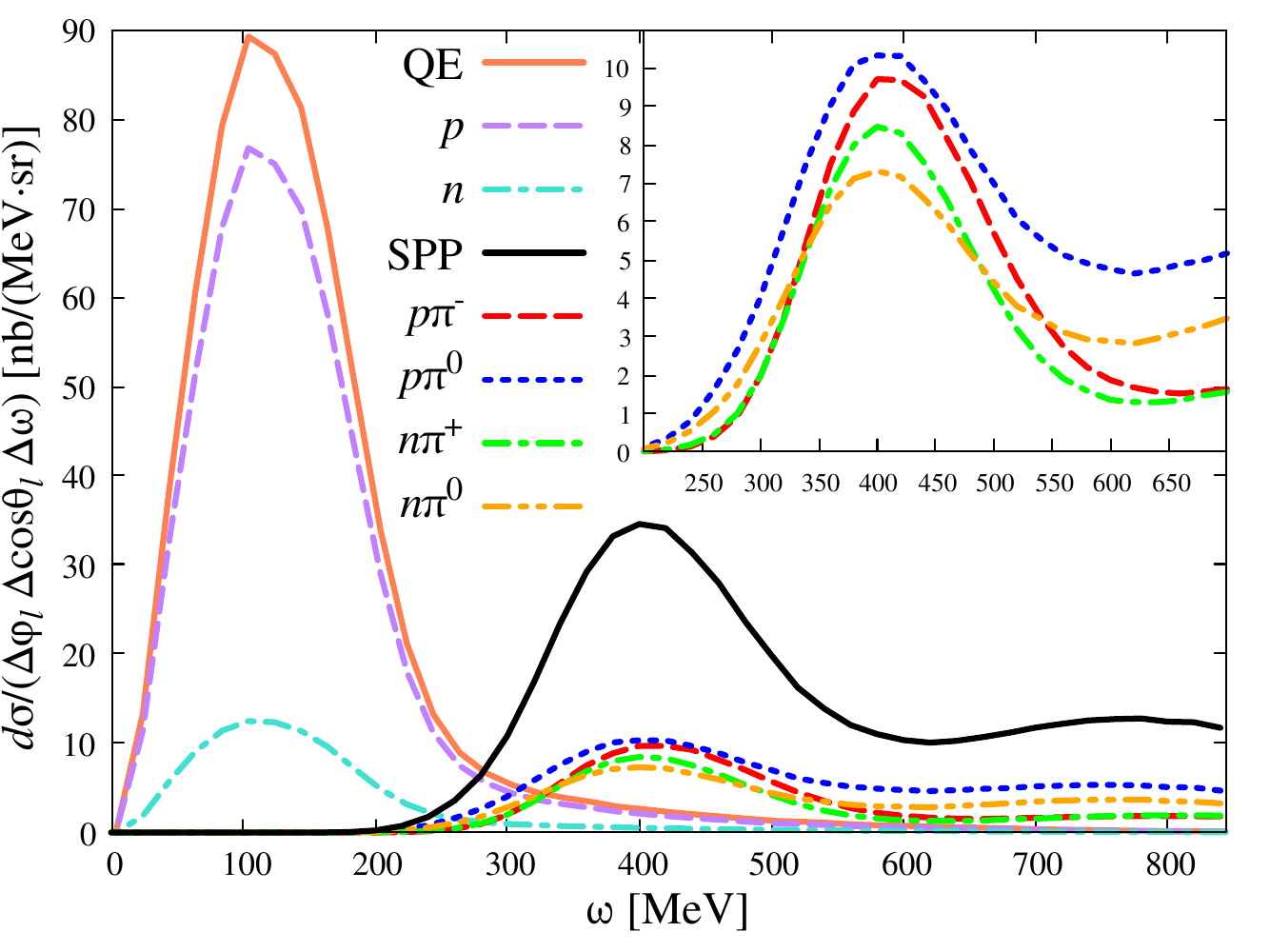}
\caption{Model prediction for the inclusive electron scattering cross section for $E_i=855$ MeV, $\theta_l=(30.00\pm1.15)$ deg and $\phi_l=(180\pm4)$ deg. 
The inner plot is a zoom around the Delta peak focusing on the SPP channels.}
\label{fig:MAMI_inclusive}
\end{figure}

In this work, we consider a beam energy of $E_i=855$ MeV.
The electron scattering cross section grows rapidly for smaller scattering angles (see inclusive electron scattering datasets at Ref.~\cite{ee-data}), 
on the other hand, at very forward angles, QE and other low-energy processes dominate, with little contribution from pion production.
Hence, we have chosen $\theta_l=30$ deg, so the conditions explained above are relatively well met. 
The inclusive cross section in terms of energy transfer, $\omega$, for all reaction channels considered in our simulations, is shown in Fig.~\ref{fig:MAMI_inclusive}. The model for the QE contribution is the one presented in Ref.~\cite{Franco-Munoz23}, with a one-body current operator and the ED-RMF potential for final states.

\subsubsection{Signal and background}

\begin{figure*}
\centering
\includegraphics[width=.49\textwidth,angle=0]{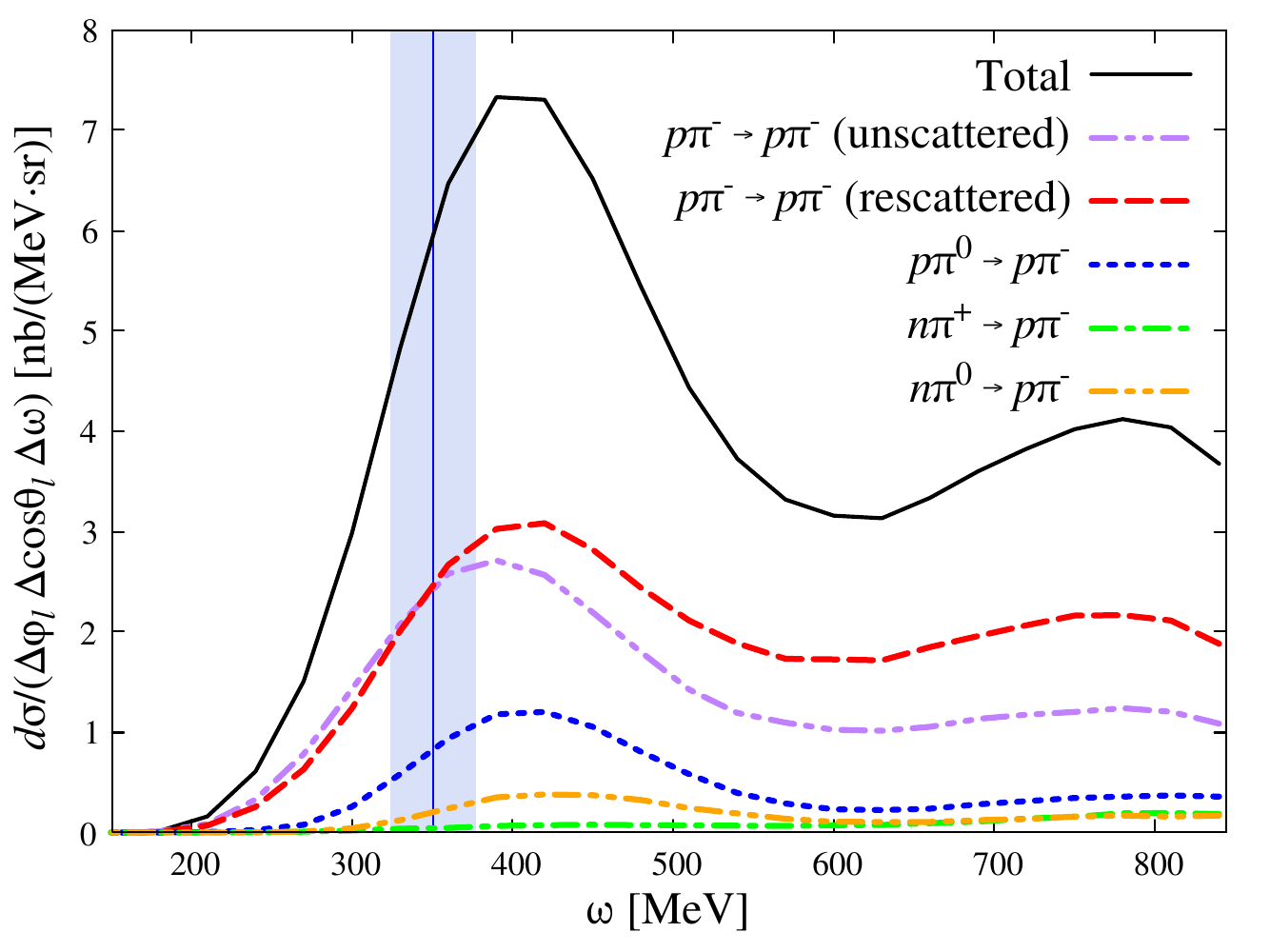}
\includegraphics[width=.49\textwidth,angle=0]{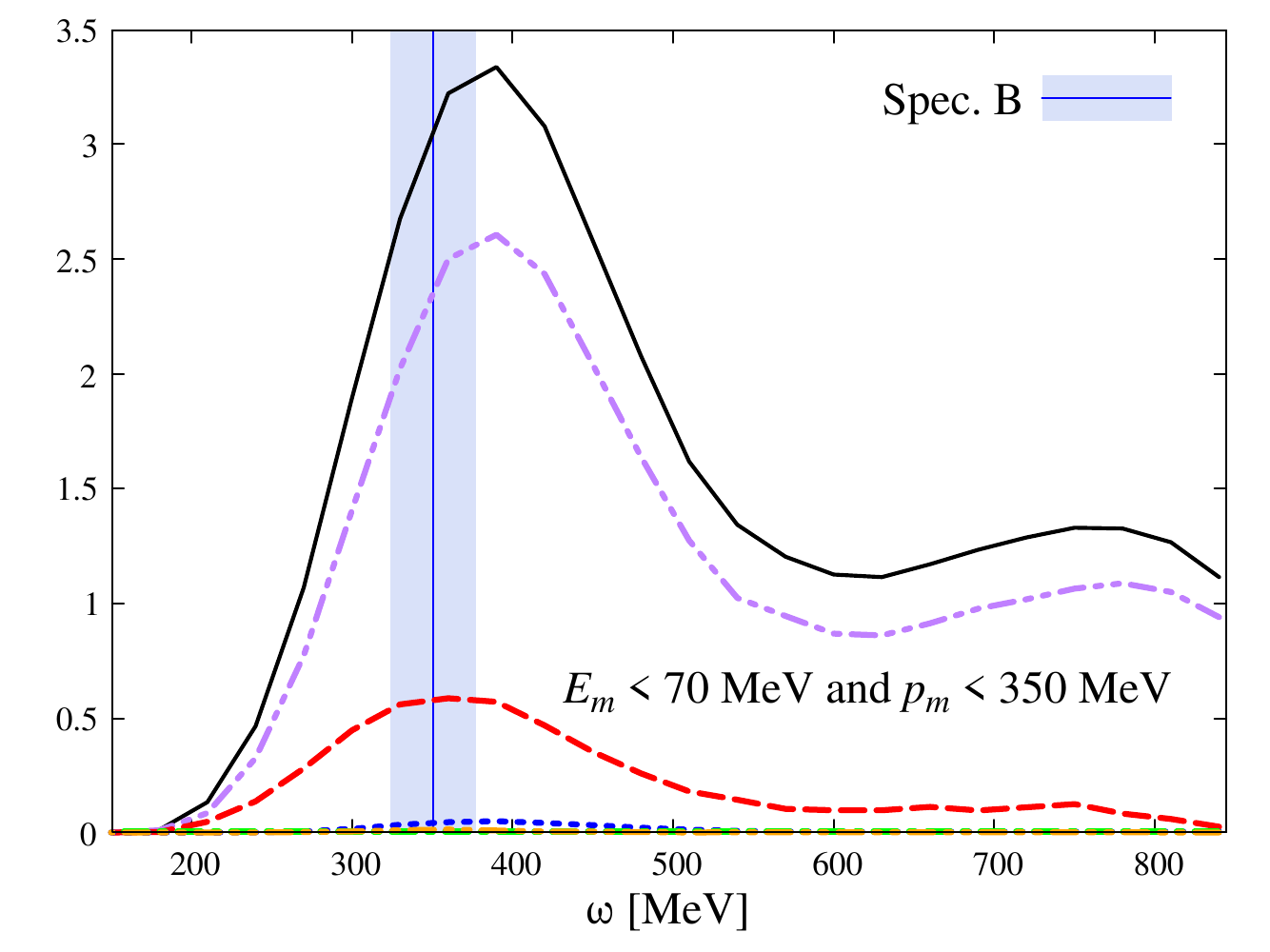}
\caption{The left panel shows the cross sections corresponding to events with at least one proton and one $\pi^-$ in the final state, arose from different reaction channels: $p\pi^-$, $p\pi^0$, $n\pi^+$ and $n\pi^0$. The events from the $p\pi^-$ channel are separated in unscattered and rescattered  contributions. The right panel is the same, but only events with $E_m<70$ MeV and $p_m<350$ MeV contribute. The blue line indicates the central energy chosen for the final electron, the band represents the acceptance of Spec. B. In all cases, $E_i=855$ MeV, $\theta_l=(30\pm1.15)$ deg and $\phi_l=(180\pm4)$ deg.}
\label{fig:atleast-1p1pi-}
\end{figure*}

To account for possible rescatterings, the hadrons from the events that contribute to the cross sections in Fig.~\ref{fig:MAMI_inclusive} were propagated in the \textsc{NuWro} INC. The cross sections resulting from the selection of events with ``at least one proton and one negative pion in the final state'' (in what follows we refer to it as the $1p1\pi^-$ sample) are presented in Fig.~\ref{fig:atleast-1p1pi-}.
By ``unscattered'' we refer to the events arose in the $p\pi^-$ channel~\eqref{signal} that did not suffer any rescattering. This is the desired signal. In the left panel, we see that it is approximately 35\% of the $1p1\pi^-$ sample around the Delta peak.
The dominant background (around 40\%) corresponds to events that originated in the $p\pi^-$ channel~\eqref{signal} but suffered rescattering.  
Events from the reactions 
\ba
  e+p\longrightarrow e'+p+\pi^0\,,\\
  e+n\longrightarrow e'+n+\pi^0,
\ea
that were subject to CEX contribute significantly to the 1p1$\pi^-$ sample, with approximately 16\% and 5\%, respectively. 

In the right panel in Fig.~\ref{fig:atleast-1p1pi-}, we show the same as in the left panel but after applying cuts in missing energy and momentum, specifically, only events with $E_m<70$ MeV and $p_m<350$ MeV contribute. With these cuts, we eliminate most of the background. The cross section is now approximately half of the original one, but the unscattered events contribute more than 75\% of the full $1p1\pi^-$ cross section around the Delta peak.

The energy of the final electron is selected so we are close to the Delta-resonance peak, i.e. maximum of SPP cross section, but below it, so the contribution from two-pion production and higher-mass resonances is negligible. This considerably simplifies the theoretical modeling of the SPP process.
The vertical blue line and band in Fig.~\ref{fig:atleast-1p1pi-} represent the final electron energy chosen and the acceptance of spectrometer B.

The contribution of QE events to the $1p1\pi^-$ sample was evaluated and found to be smaller than 1\% even before the cuts in $E_m$ and $p_m$.

\subsubsection{Hadron spectrometers}
\begin{figure*}[ht]
\centering  
\includegraphics[width=.49\textwidth,angle=0]{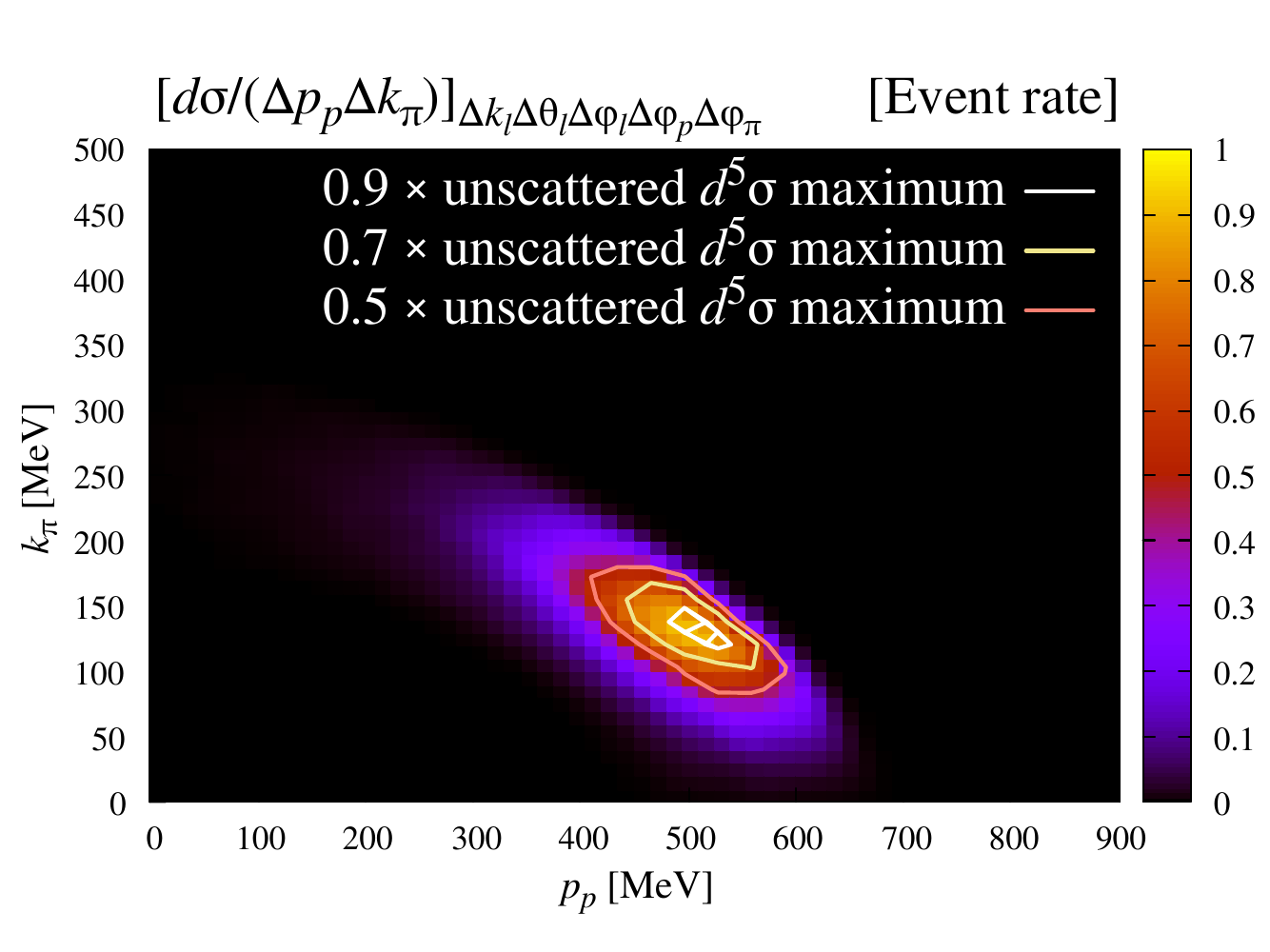}
\includegraphics[width=.49\textwidth,angle=0]{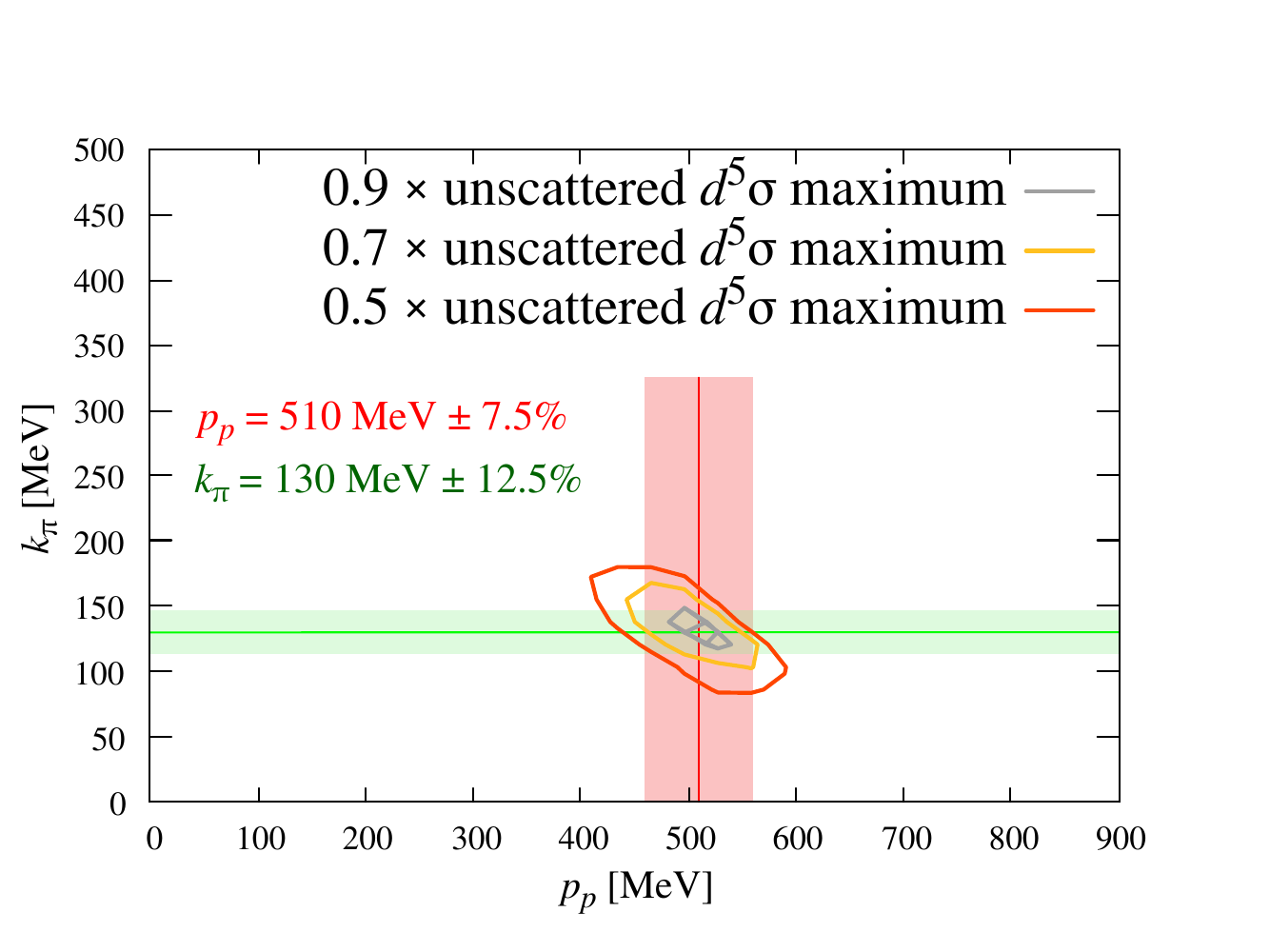}
\includegraphics[width=.49\textwidth,angle=0]{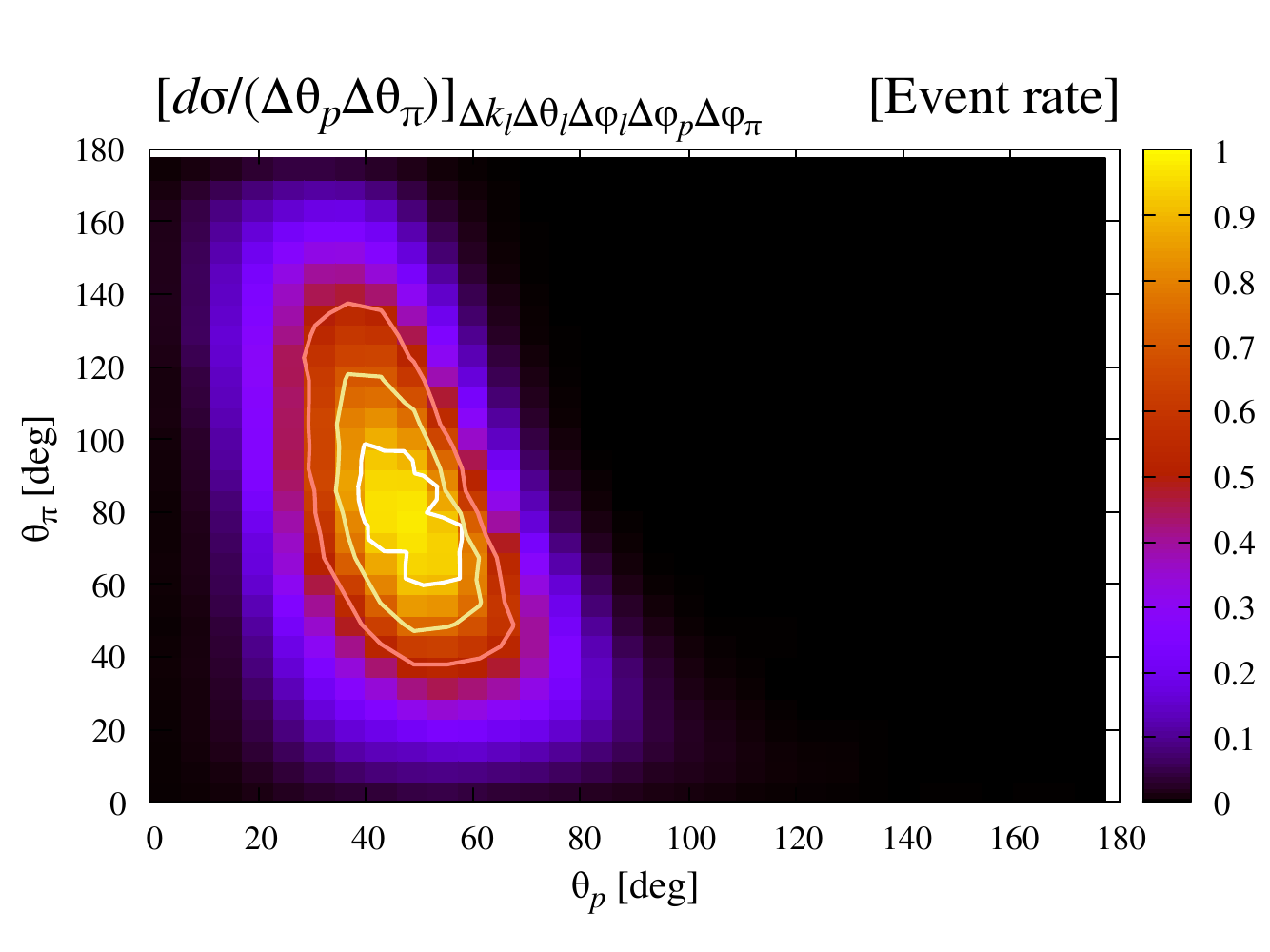}
\includegraphics[width=.49\textwidth,angle=0]{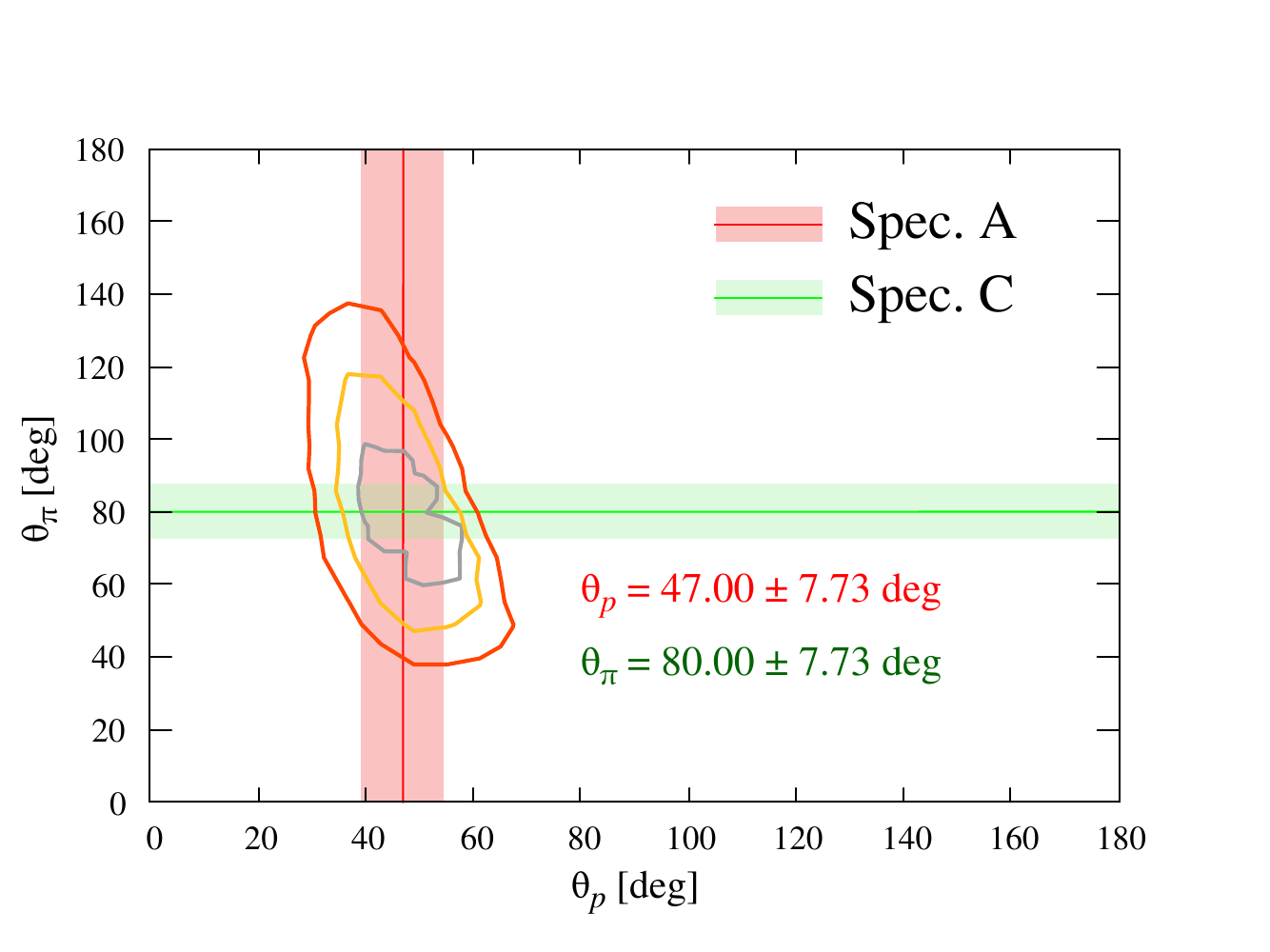}
\caption{Left panels: Differential cross sections in terms of different combinations of hadron variables. Right panels: Coverage of the proton (red, Spec. A) and pion (green, Spec. C) spectrometers, the line indicates the central value and the band represents the acceptance. 
Upper row: proton and pion momenta. Lower row: scattering angles of the proton and pion. 
The line contours indicate the regions where the unscattered $1p1\pi^-$ cross section are 90\%, 70\% and 50\% of its maximum. %
Only events with $E_m<70$ MeV and $p_m<350$ MeV contribute. 
In all cases, $E_i=855$ MeV, $E_l=505 \text{ MeV} \pm7.5\%$, $\theta_l=(30\pm1.15)$ deg, $\phi_l=(180\pm4)$ deg, $\phi_p=(0\pm4)$ deg and  $\phi_\pi=(180\pm4)$ deg.
}
\label{fig:where-to-look}
\end{figure*}

The goal is to have large cross sections while minimizing the background as much as possible. Next step is to select the energy and scattering angle for the final hadrons. Thus, we look at the differential cross sections for fixed lepton kinematics (same as in Fig.~\ref{fig:MAMI_inclusive}), and place the final proton and pion on the plane:
\begin{equation}
    \phi_p=(0\pm4)\text{ deg}\quad,\quad\phi_\pi=(180\pm4)\text{ deg.}
\end{equation}
Here, we have taken into account the restrictions summarized in Table~\ref{table:resolution}.
In the left panels of Fig.~\ref{fig:where-to-look}, we show the differential cross sections
\begin{itemize}
    \item In terms of pion and proton momenta, first row: $d\sigma/(\Delta p_p\Delta k_\pi)$,
    \item In terms of pion and proton scattering angle, second row: $d\sigma/(\Delta\theta_p\Delta\theta_\pi)$,
\end{itemize}
both computed within the restricted phase space $(\Delta\Omega_l\Delta k_l\Delta\phi_p\Delta\phi_\pi)$. Superimposed on the cross-section plots we show contours corresponding to the regions where the unscattered $1p1\pi^-$ cross section is 90\%, 70\% and 50\% of its maximum. This shows that the maximum of the unscattered cross section mainly coincides with that of the $1p1\pi^-$ sample, where rescattering is present. 
This result is due to the kinematic restrictions and, essentially, model independent. 

In the right panels of Fig.~\ref{fig:where-to-look}, we show the selected central values for the proton (red, Spec. A) and pion (green, Spec. B) momenta; the bands represent the ranges covered by the detector acceptances.
The selection of pion and proton momenta is such that the ratio signal to background is maximum; however, as shown previously (left panels of Fig.~\ref{fig:where-to-look}), this phase-space region coincides with the one in which the cross section is maximum.

\begin{figure}[ht]
\centering  
\includegraphics[width=.49\textwidth,angle=0]{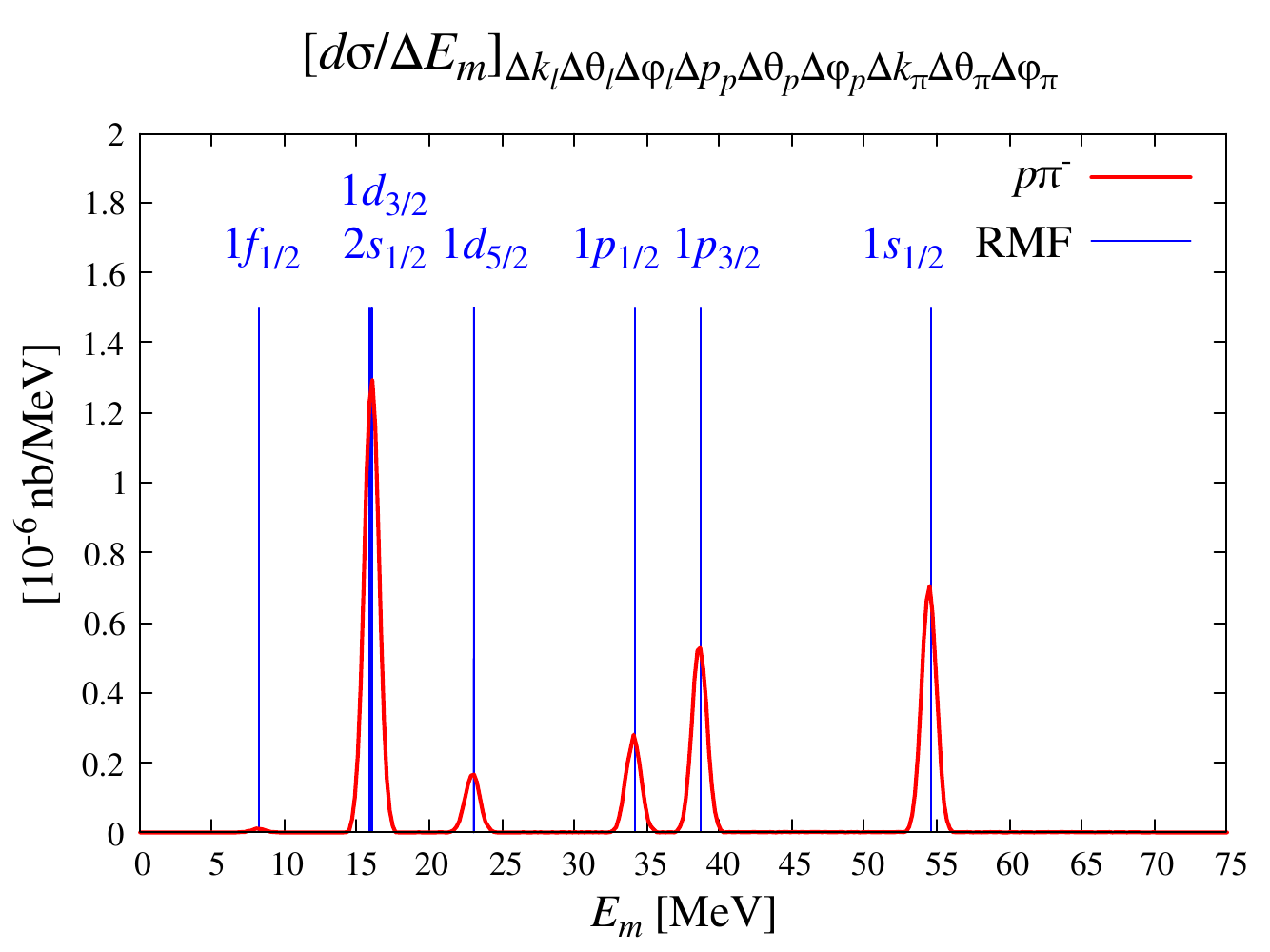}
\caption{Differential cross section in terms of missing energy for the signal sample $1p1\pi^-$ within the chosen acceptances for MAMI. Vertical lines represent the RMF eigenvalues for the shells. Only events with $E_m<70$ MeV and $p_m<350$ MeV contribute. }
\label{fig:MAMI_Em}
\end{figure}

The cross section in terms of missing energy is shown in Fig.~\ref{fig:MAMI_Em}. We see that the shape corresponds to that described in Sec.~\ref{theomodels}: Gaussians with $\sigma=0.5$ MeV centered in the eigenvalues given by the RMF. 

\begin{figure}[ht]
\centering  
\includegraphics[width=.49\textwidth,angle=0]{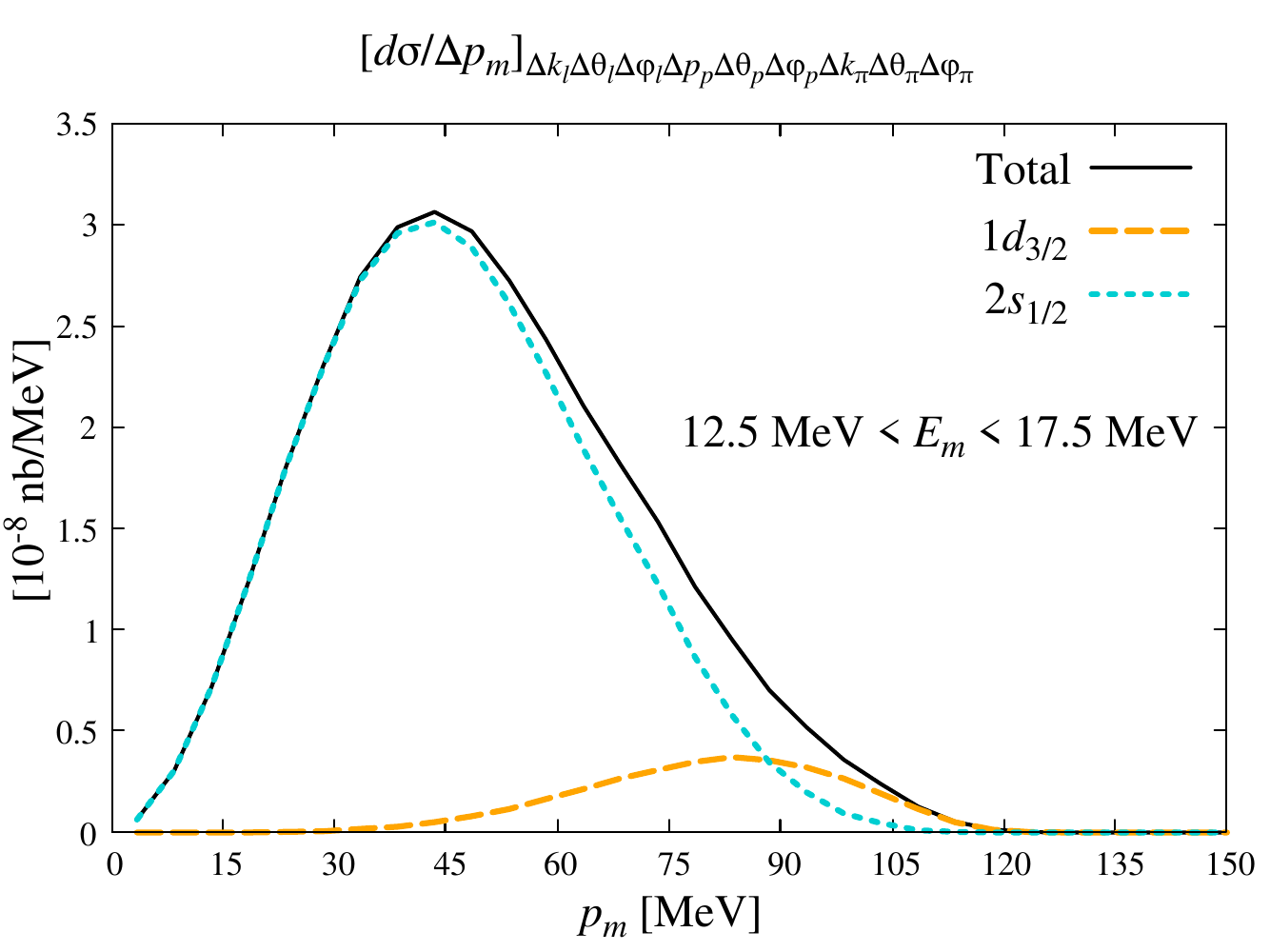}
\caption{Differential cross sections in terms of missing momentum for $12.5<E_m<17.5$ MeV. The turquoise dotted (orange dashed) line correspond to the $2s_{1/2}$ ($1d_{3/2}$) shell. Only events with $E_m<70$ MeV and $p_m<350$ MeV contribute.}
\label{fig:MAMI_pm}
\end{figure}

Furthermore, due to the small acceptances of the spectrometers, any background is gone, i.e., all the events are unscattered. This can be understood as rescattering causing angles and energies to deviate, and hence, rescattered hadrons will not reach the detector.
This shows that measuring this cross section, one could, in principle, get access to the energy levels of neutrons, spectroscopic factors, or more generally, to the mean-field part of the $^{40}$Ar spectral function.
Each peak corresponds to a different energy level (see Table~\ref{table:levels}), except for the levels $1d_{3/2}$ and $2s_{1/2}$ which overlap. 
The momentum distributions of these two $l=0$ and $l=2$ states are different, so one could try to disentangle them by looking at the cross section in terms of $p_m$ in the corresponding $E_m$ window.
This is shown in Fig.~\ref{fig:MAMI_pm}, one sees that the $l=0$ is narrower and has most of the strength below $p_m=100$ MeV, while the $l=2$ is broader and shifted towards larger $p_m$.
We emphasize that, given the excellent resolution of the MAMI spectrometers, see Table~\ref{table:resolution}, it is not necessary to include any smearing in the simulation of the detected particles.

\begin{figure}[h]
\centering
\includegraphics[width=.5\textwidth,angle=0]{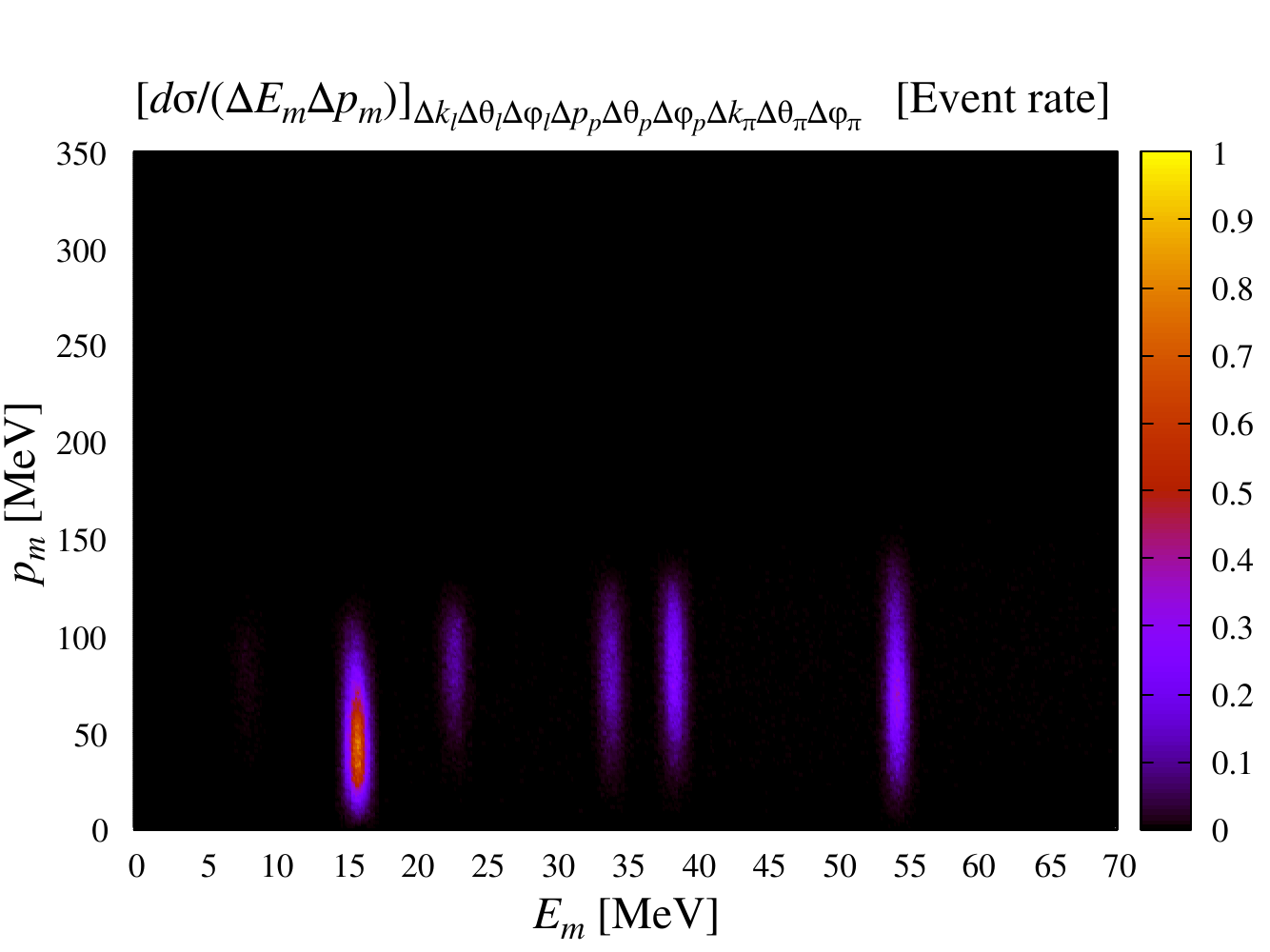}
\caption{Differential cross section in terms of $E_m$ and $p_m$ for the chosen acceptances at MAMI. Only events with $E_m<70$ MeV and $p_m<350$ MeV contribute. }
\label{fig:Em-pm_mami}
\end{figure}

For completeness, in Fig.~\ref{fig:Em-pm_mami} we include the differential cross section in terms of $E_m$ and $p_m$. 

\subsubsection{Event rate}

It is interesting to estimate the amount of events for a given run time that is expected, according to our cross section prediction. 
Given our cross section prediction for this triple coincidence experiment, together with the acceptances of the spectrometers and the $E_m$-$p_m$ cuts, a sufficiently dense target would be needed for the feasibility of the experiment. 

One could consider a liquid argon target, in which case, following Ref.~\cite{MAMI_Ar40_Proposal_2017}, luminosities of 
$4\times10^7$ ($\mu$b$\cdot$s)$^{-1}$ could be reached.
In that case, the event rate is
\begin{equation}\label{Eq:eventrate}
N = \mathcal{L}\cdot\sigma\cdot{\varepsilon}\approx 0.002\text{ events}/s,
\end{equation}
where we have assumed perfect efficiency, $\varepsilon=1$, and $\sigma\approx3.8\times10^{-6}$ nb, which is the integral of the differential cross section in Fig.~\ref{fig:MAMI_Em}. 
Larger cross sections could be achieved by reducing the beam energy and the lepton scattering angle (up to 7 deg, the minimum possible angle for spectrometer B).

\subsubsection{Out-of-plane capability}

\begin{figure}[ht]
\centering  
\includegraphics[width=.49\textwidth,angle=0]{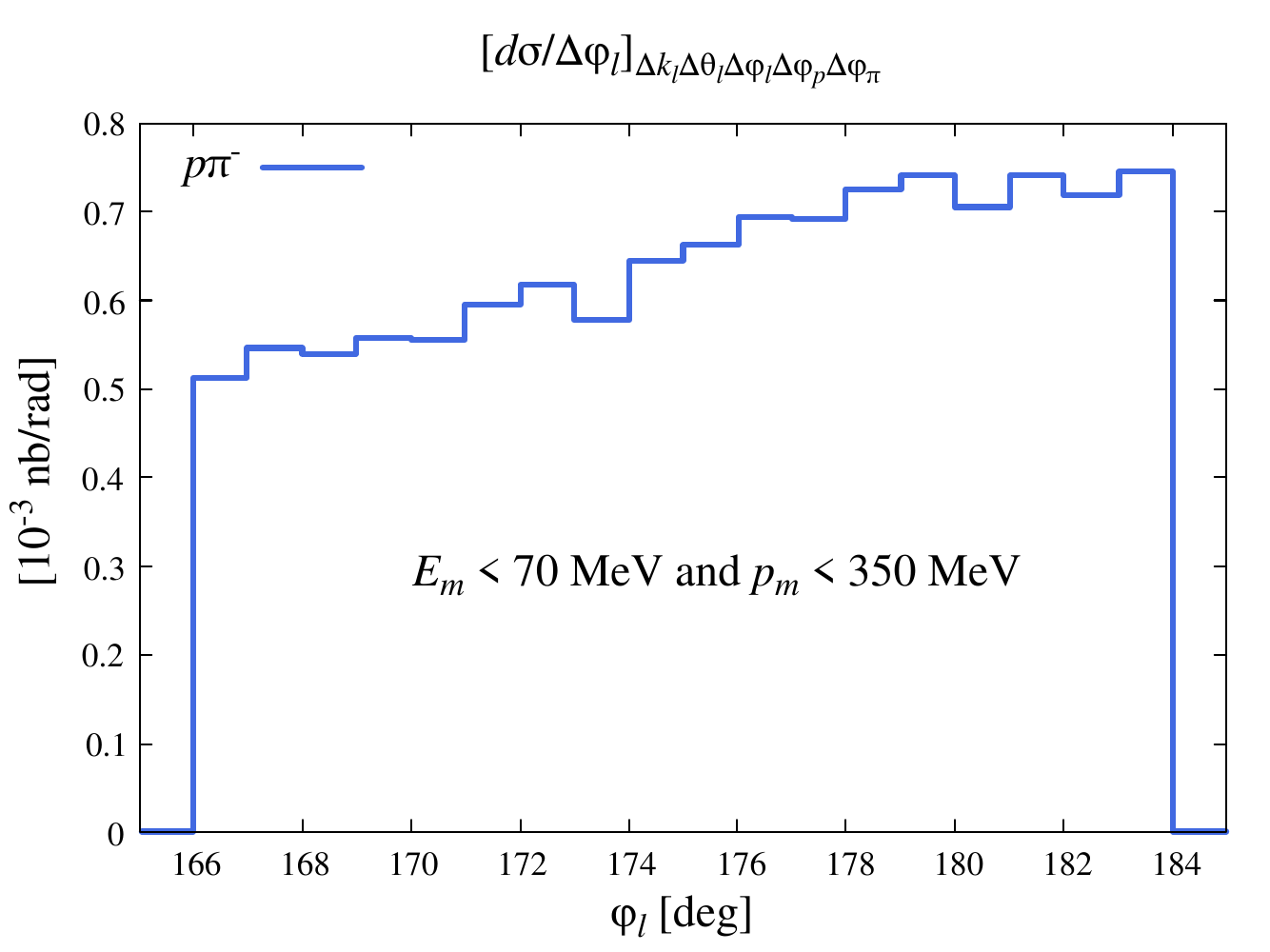}
\caption{Differential cross section in terms of $\phi_l\in\{170\text{ deg},180\text{ deg}\}\pm 4$ deg. The calculation has been done for fixed $\phi_p=(0\pm4)$ deg and $\phi_\pi=(180\pm4)$ deg. Also, $E_i=855$ MeV, $E_l=505$ MeV $\pm 7.5\%$ and $\theta_l=(30\pm15)$ deg. Each step is 1 deg wide.}
\label{fig:out-of-plane}
\end{figure}

As noted in Table~\ref{table:resolution}, Spec. B has the capability of going out-of-plane by 10 deg. To round out the discussion, we compute the cross section for $\phi_l\in \{170\text{ deg},180\text{ deg}\}\pm 4$ deg. We also keep the proton and the pion on plane, and apply cuts in $E_m$ and $p_m$. The result shown in Fig.~\ref{fig:out-of-plane} confirms that the cross section is maximum at $\phi_l\sim180$ deg. 

\subsection{CLAS}

CLAS is a large solid angle spectrometer at the Thomas Jefferson National Accelerator Facility for electron beams~\cite{CLASwebsite}. 

\begin{figure}[ht]
\centering
\includegraphics[width=.49\textwidth,angle=0]{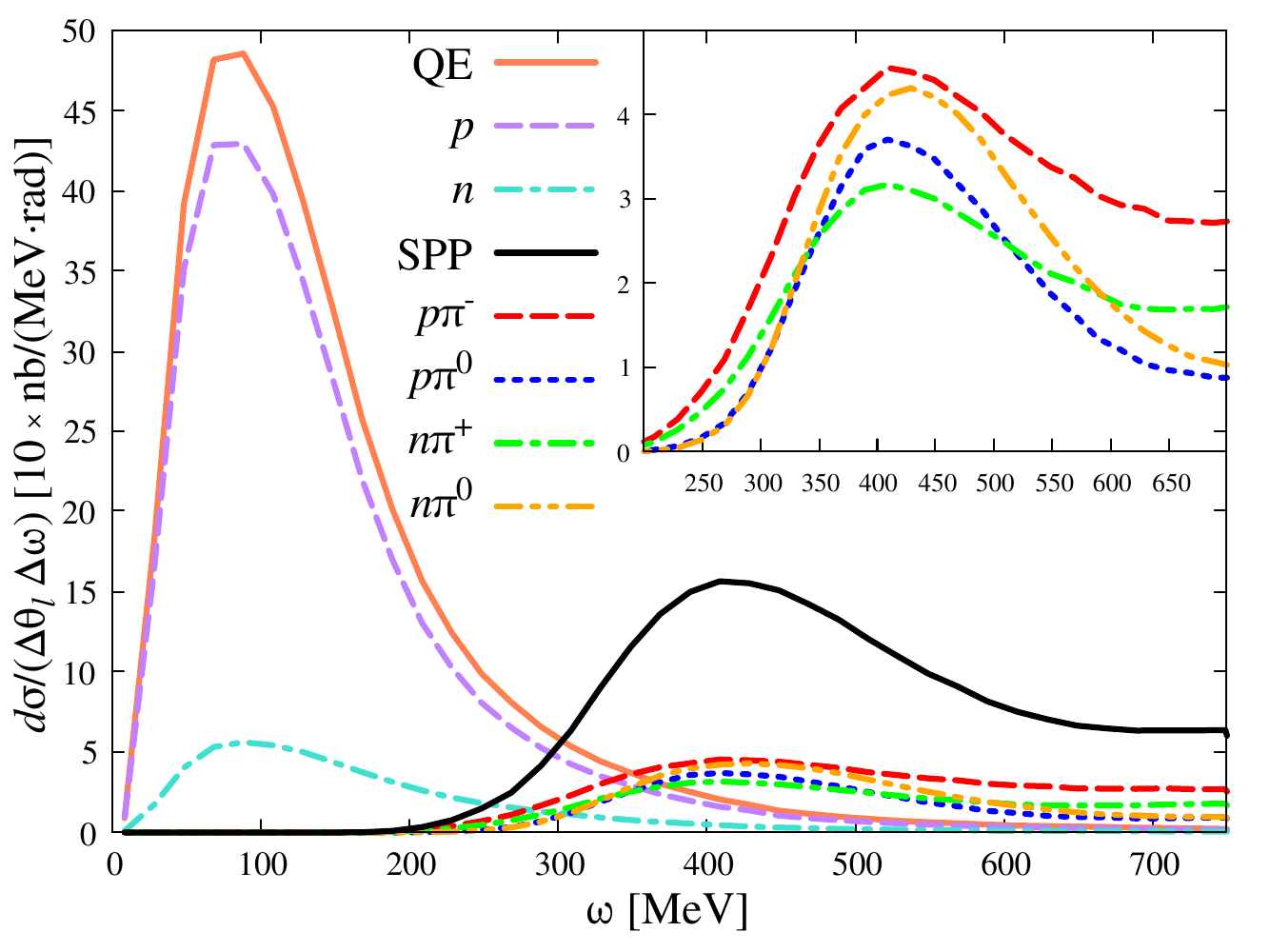}
\caption{Model prediction for the CLAS inclusive electron scattering cross section with $E_i=1159$ MeV and $\theta_l=(30\pm15)$ deg. 
The inner plot is a zoom in around the Delta peak with the SPP channels only.}
\label{fig:CLAS_inclusive_nofsi}
\end{figure}

We perform the same analysis as in the MAMI simulation (and used the same theoretical models). The semi-inclusive cross section for these reaction channels using the CLAS beam energy and lepton scattering angle acceptance is shown in Fig.~\ref{fig:CLAS_inclusive_nofsi}. 
We use $E_i=1159$ MeV and the same acceptances, energy thresholds and momentum resolution as in~\cite{Khachatryan21}. The details are listed in Table \ref{table:CLAS}.

        \begin{table}
        \begin{tabular}{c l c c c c c c c c c c c c c l c }
        \hline
        \hline
        \arrayrulecolor{white}\hline
        \hline
        & Particle & & & & $p$ [MeV] & & & & $\theta$ [deg] & & & & Mom. resolution &\\
        \arrayrulecolor{white}\hline
        \hline
        \arrayrulecolor{black}\hline
        \arrayrulecolor{white}\hline
        \hline
        & Proton & & & & $>$ 300 & & & &  $>$ 10 & & & & 3.0\%\\ 
        & Lepton & & & & $>$ 400 & & & & 30 $\pm$ 15  & & & & 1.5\%\\  
        & Pion & & & & $>$ 150   & & & & $>$ 22 & & & & 2.1\%\\
        \arrayrulecolor{black}\hline
        \hline
        \end{tabular}
        \caption{Acceptances, energy thresholds and momentum resolutions used in our simulation for CLAS (values taken from~\cite{Khachatryan21}).}
        \label{table:CLAS}
        \end{table}

\begin{figure*}[ht]
\centering  
\includegraphics[width=.49\textwidth,angle=0]{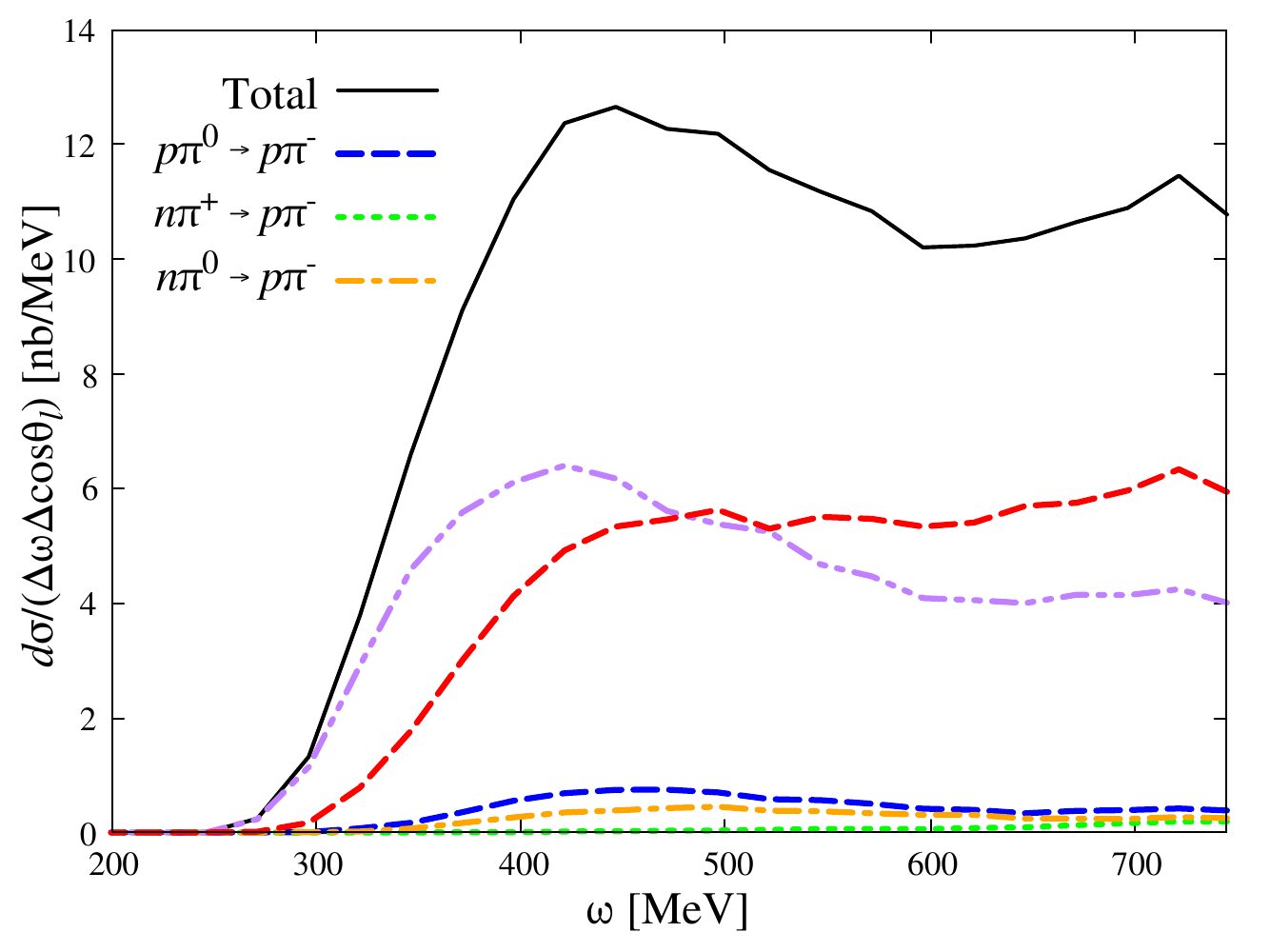}
\includegraphics[width=.49\textwidth,angle=0]{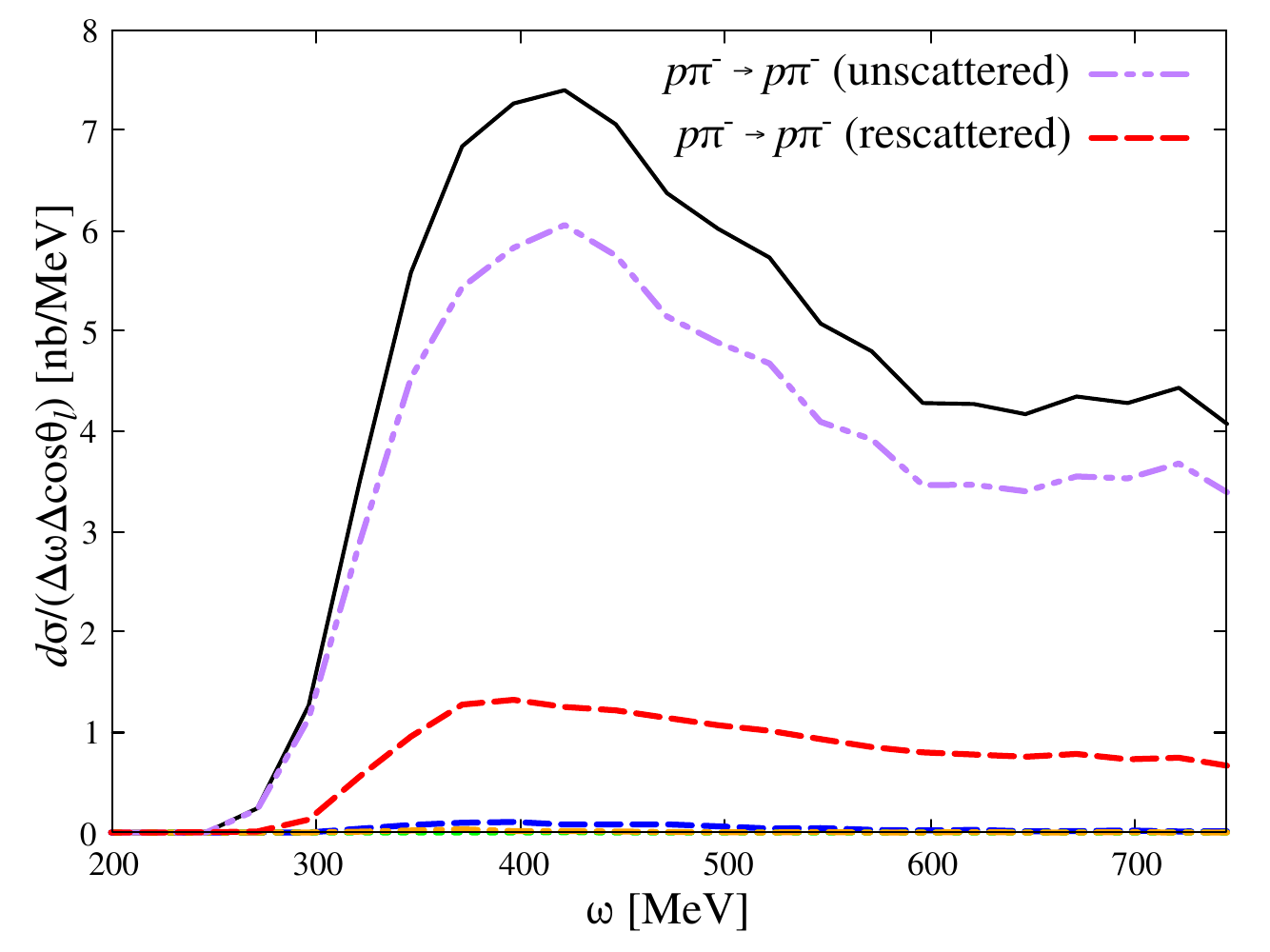}
\caption{The left panel shows the cross sections corresponding to events with at least one proton and one $\pi^-$ in the final state, arose from different reaction channels: $p\pi^-$, $p\pi^0$, $n\pi^+$ and $n\pi^0$. The events from the $p\pi^-$ channel are separated in unscattered and rescattered  contributions. The right panel is the same, but only events with $E_m<70$ MeV and $p_m<350$ MeV contribute. The kinematical conditions are specified in Table~\ref{table:CLAS}. Smearing of the final particles' momenta is applied. In all cases, $E_i=1159$ MeV, $\theta_l=(30\pm15)$ deg and $k_l<400$ MeV. 
}
\label{fig:CLAS_inclusive}
\end{figure*}

The predicted cross section contributing to the $1p1\pi^-$ sample are shown in Fig.~\ref{fig:CLAS_inclusive}  as a function of the energy transfer.
In the right panel, we applied the same $E_m\text{-}p_m$ cuts as in the MAMI simulation. These cuts produce the expected effect of eliminating most of the background. 

\begin{figure}[ht]
\centering  
\includegraphics[width=.49\textwidth,angle=0]{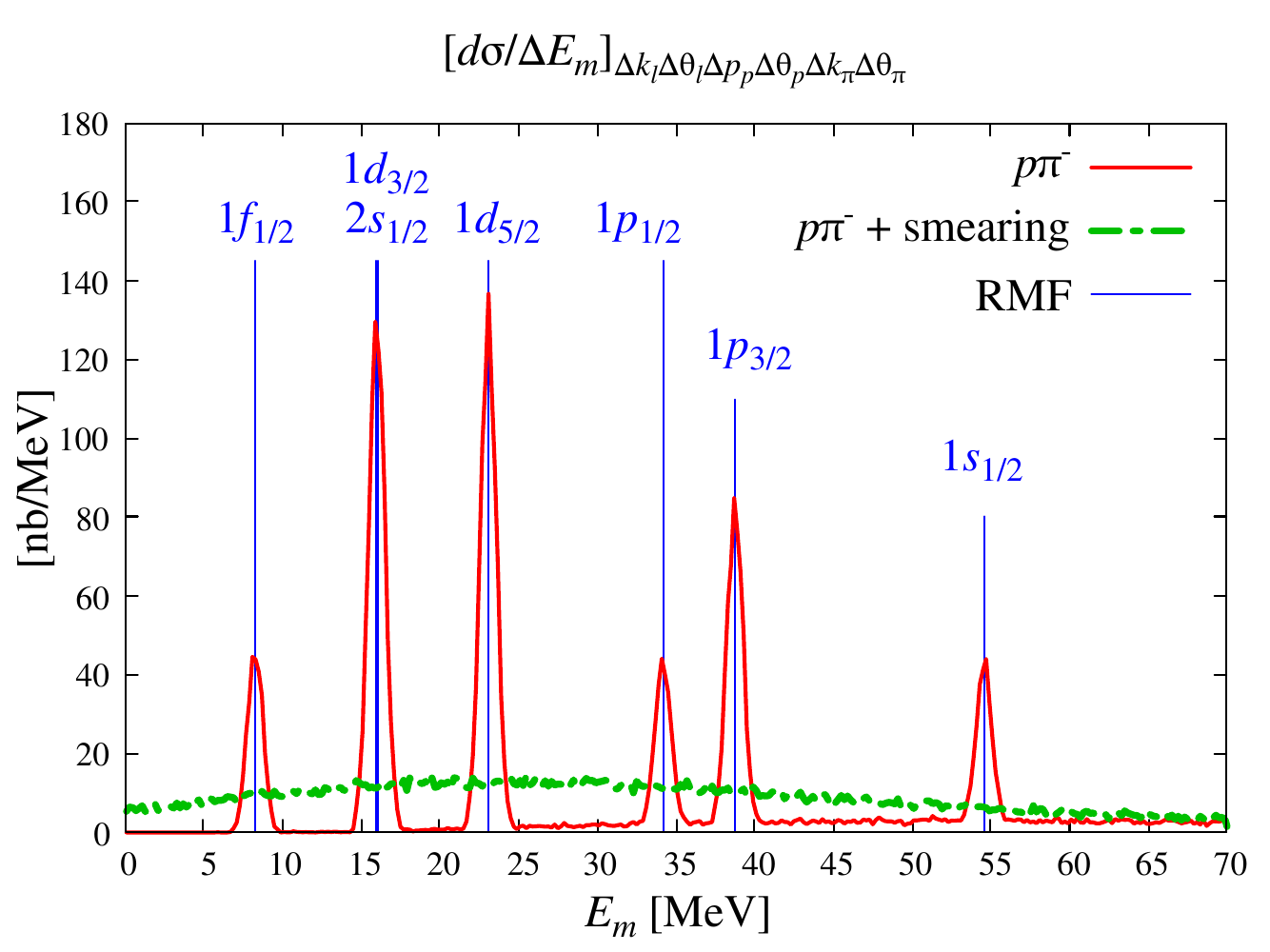}
\caption{Differential cross section in terms of missing energy for the signal sample $1p1\pi^-$ within the given acceptances for CLAS. We show the distributions with rescattering (red), and with rescattering and the momentum smearing (dashed-dotted green)Vertical. Lines represent the RMF eigenvalues for the shells. Only events with $E_m<70$ MeV and $p_m<350$ MeV contribute.}
\label{fig:CLAS_Em}
\end{figure}

To simulate the energy resolution of the CLAS detector, we smear the momenta of the final particles using Gaussian distributions with standard deviations of 1.5\% for the lepton, 3.0\% for the proton and 2.1\% for the pion.
The effect of smearing is shown in Fig.~\ref{fig:CLAS_Em}. It is clear that the shell structure cannot be resolved due to the energy resolution.

\begin{figure*}[ht]
\centering  
\includegraphics[width=.32\textwidth,angle=0]{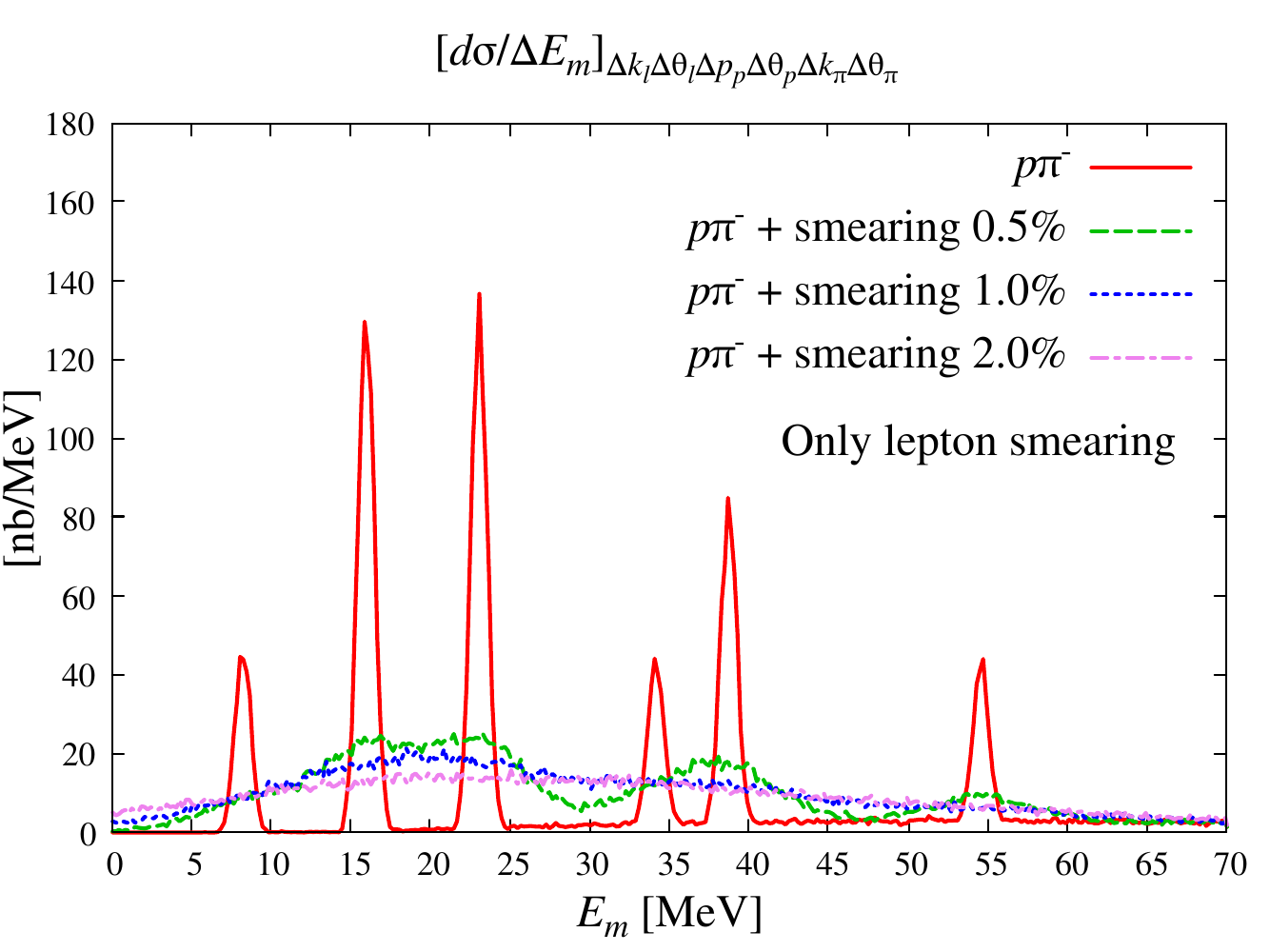}
\includegraphics[width=.32\textwidth,angle=0]{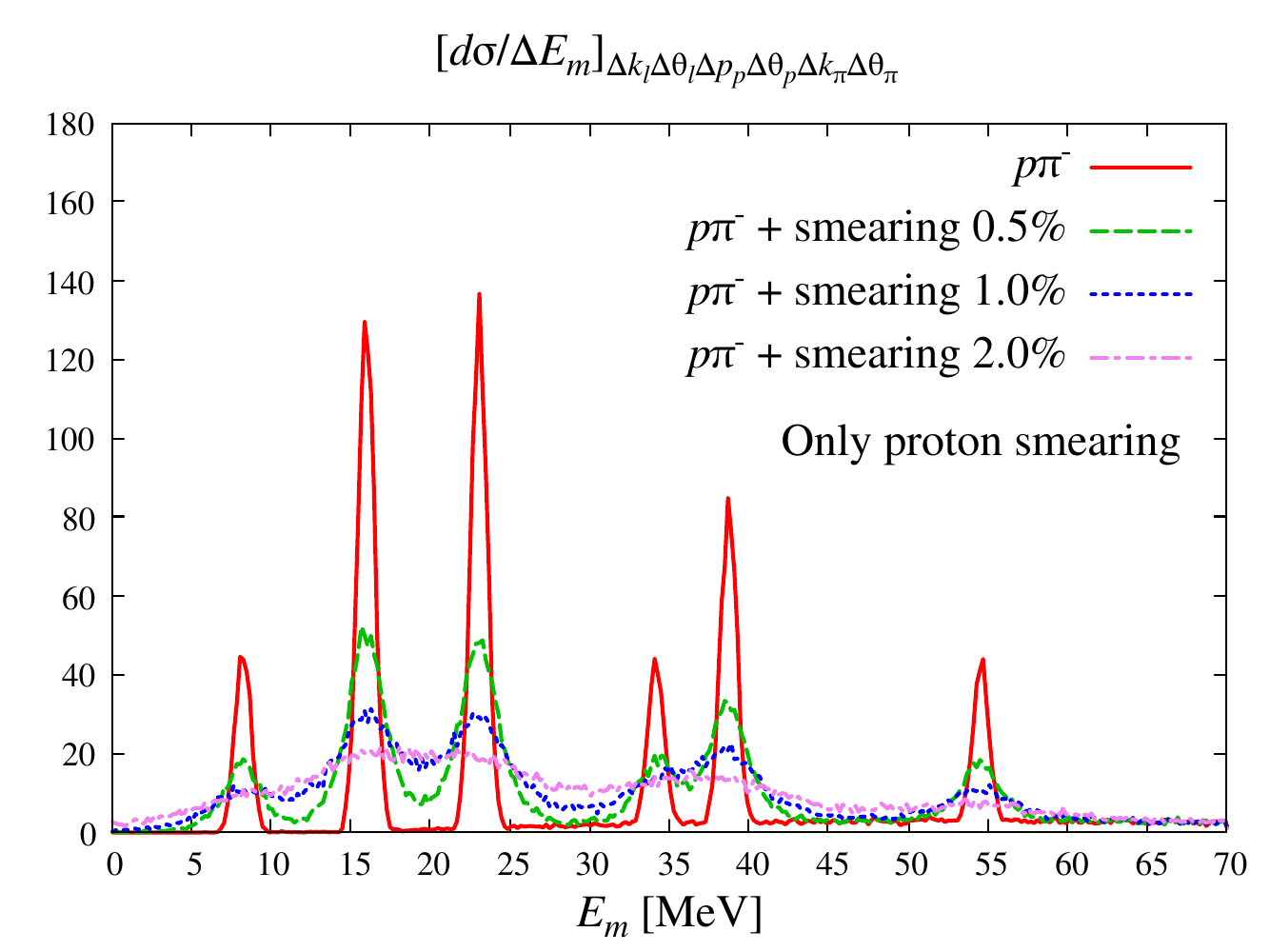}
\includegraphics[width=.32\textwidth,angle=0]{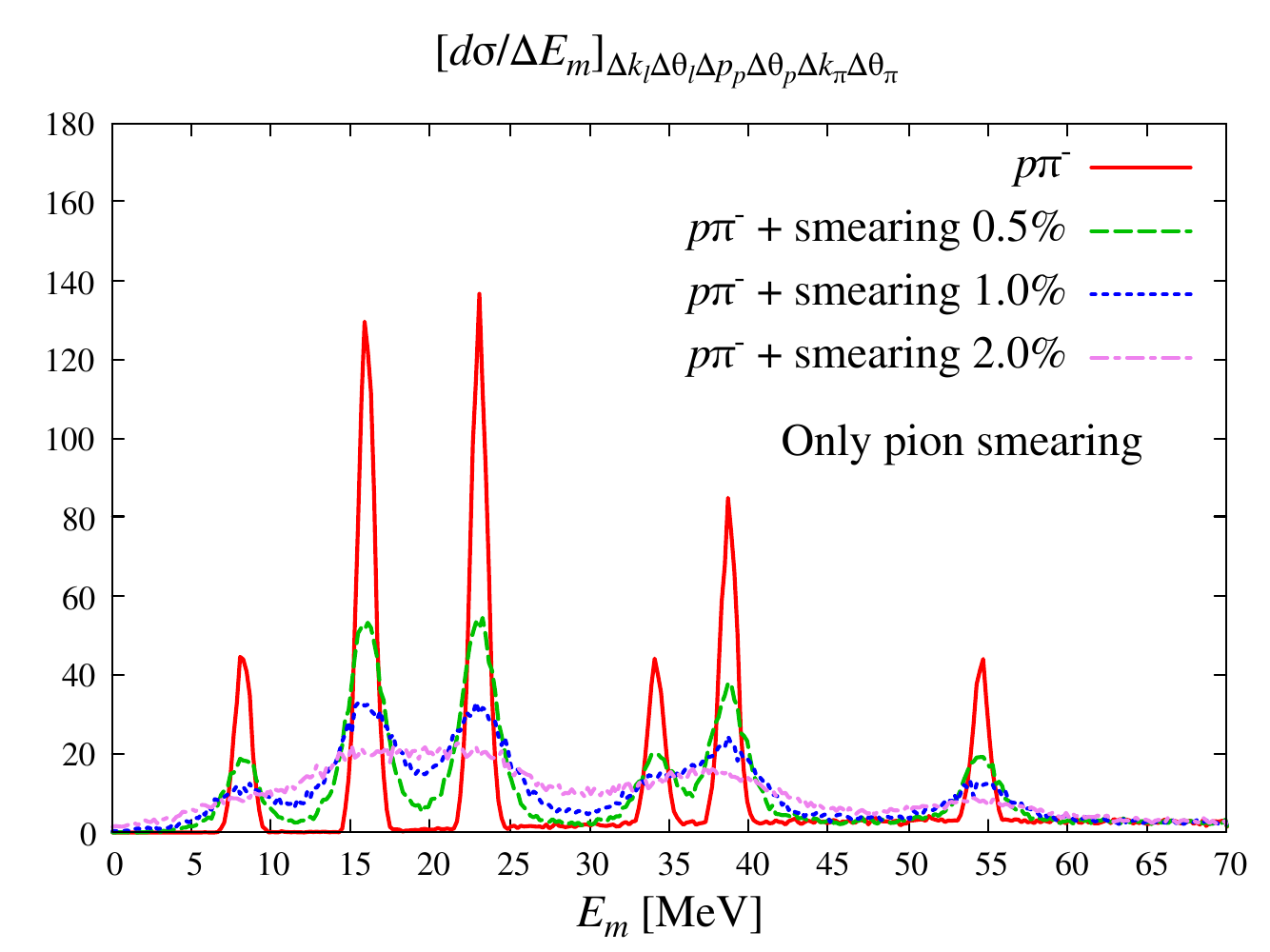}
\caption{Study of momentum resolution in CLAS. Panels show the differential cross section in terms of missing energy for four cases: no smearing and momentum smearing with $\sigma=0.5\%$, 1.0\% and 2.0\%. In each panel, the smearing is applied only to the momentum of one detected particle. From left to right: lepton, proton, pion. Only events with $E_m<70$ MeV and $p_m<350$ MeV contribute} 
\label{fig:CLAS_resolution}
\end{figure*}

In Fig.~\ref{fig:CLAS_resolution}, we show the effect of the smearing in the momenta of the particles. We show the same distribution as in Fig.~\ref{fig:CLAS_Em}, but the smearing is {\it only} applied to the lepton (left), proton (middle) or pion (right panel). The momentum resolution applied in each case is indicated in its corresponding figure. 
It can be seen that the signal is very sensitive to smearing the lepton momentum, while not that much for the proton and pion ones.

\section{Conclusions}

We propose a triple coincidence experiment $^{40}$Ar$(e,e'p\pi^-)$ as a way to constrain the neutron structure in $^{40}$Ar. 
We have performed theoretical predictions for the cross sections and estimated the main background for kinematics, acceptances and energy resolution compatible with the MAMI and CLAS facilicites. 

Our predictions are based on the RPWIA. Hadron rescatterings are accounted for by propagating the hadrons through the \textsc{NuWro} INC. 

We have shown that most of the background can be eliminated by restricting the samples to small values of missing energy, $E_m<70$ MeV, and missing momentum, $p_m<350$ MeV. This is because, due to rescattering, strength is moved from low to high $E_m\text{-}p_m$ regions.

In the case of MAMI, we have studied the optimal position for the three spectrometers. The electron spectrometer is chosen so that the SPP cross section is maximum compared to the contribution from other channels, this is done by studying the inclusive cross section.
The hadron spectrometers are chosen so that the unscattered channel (events from the $p\pi^-$ channel~\eqref{signal} in which the $\pi^-$ and the $p$ do not suffer any rescattering) is maximum. This region coincides with the region where the $1p1\pi^-$ cross section is maximum. This can by explained by the fact that the rescattering background is completely negligible with a spectrometer setup that tightly centers around the signal region. Moreover, we observe that, for a triple-coincidence experiment such as the one considered in this work, a liquid argon target is essential to achieve sufficiently high statistics.
In the case of CLAS, we have performed a similar study but taking into account the experimental conditions of this facility. 
The larger acceptances of CLAS allow for larger cross sections.
We have seen that most of the background can be removed by applying cuts in $E_m$ and $p_m$. Unfortunately, due to the momentum resolution in the detected particles, the shell structure is completely blurred. Thus, the capabilities of the CLAS detector are better suited for other studies, such as studying nuclear effects in electron-induced single-pion production, which will be the subject of future work.\\

As we have pointed out throughout this work, this study is meant to be {\it a proof of concept}, so it opens the door to future projects. 

From a theoretical point of view, the next step would be the inclusion of the distortion of the final nucleon and pion, which are described by plane waves in this work. Moreover, regarding the evaluation of theoretical uncertainties, we point out the following:
\begin{itemize}
    \item[{\it i)}] The usage of different state-of-the-art SPP models.
    \item[{\it ii)}] Once the proton and pion distortions are included, the usage of different potentials for both hadrons.
    \item[{\it iii)}] In the case of CLAS, the usage of different INCs.
\end{itemize}

Even though several improvements can be done from the modeling perspective, one should have in mind that the conclusions obtained in this work, in particular in the case of MAMI, heavily rely on kinematics. Therefore, they are essentially model independent. 

To conclude, we would like to point out that measurements of the triple coincidence interaction can improve our understanding of nuclear effects in SPP in the Delta region, which is of interest for present and future accelerator-based neutrino experiments.\\

\begin{acknowledgments}
We thank L. Doria and J. Tena-Vidal for kindly answering all our questions on MAMI and CLAS. 
This work was supported by projects PID2021-127098NA-I00 and RYC2022-035203-I funded by MCIN/AEI/10.13039/501100011033/FEDER and FSE+, UE;
by ``Ayudas para Atracción de Investigadores con Alto Potencial-modalidad A'' funded by VII PPIT-US; and by the Fund for Scientific Research Flanders (FWO) ;
A.N. is supported by the Neutrino Theory Network (NTN) under Award Number DEAC02-07CH1135.
\end{acknowledgments}

\bibliographystyle{apsrev4-1}
\bibliography{bibliography}

\end{document}